\documentclass[reprint,showpacs,aps,pra,onecolumn,superscriptaddress]{revtex4-1}
\usepackage[T1]{fontenc}
\usepackage[latin9]{inputenc}
\usepackage{color}
\usepackage{amsmath}
\usepackage{amssymb}
\usepackage{graphicx}
\usepackage{bm}
\usepackage{epstopdf}
\usepackage[english]{babel}
\usepackage{appendix}
\usepackage{braket}
\usepackage[caption=false]{subfig}

\newcommand{\sgn}{\operatorname{sgn}}
\newcommand{\Tr}{\operatorname{Tr}}

\long\def\/*#1*/{}

\makeatletter \@ifundefined{textcolor}{} {
\definecolor{BLACK}{gray}{0}
 \definecolor{WHITE}{gray}{1}
 \definecolor{RED}{rgb}{1,0,0}
 \definecolor{GREEN}{rgb}{0,1,0}
 \definecolor{BLUE}{rgb}{0,0,1}
 \definecolor{CYAN}{cmyk}{1,0,0,0}
 \definecolor{MAGENTA}{cmyk}{0,1,0,0}
 \definecolor{YELLOW}{cmyk}{0,0,1,0}
 }
\begin{document}

\title{Driven-Dissipative Quantum Dynamics in Ultra Long-lived Dipoles in an Optical Cavity}
\author{Diego Barberena}
\affiliation{JILA, NIST, and Dept. of Physics, University of Colorado, 440 UCB, Boulder, CO  80309, USA}
\affiliation{Center for Theory of Quantum Matter, University of Colorado, Boulder, CO 80309, USA}
\author{Robert J. Lewis-Swan}
\affiliation{JILA, NIST, and Dept. of Physics, University of Colorado, 440 UCB, Boulder, CO  80309, USA}
\affiliation{Center for Theory of Quantum Matter, University of Colorado, Boulder, CO 80309, USA}
\author{James K. Thompson}
\affiliation{JILA, NIST, and Dept. of Physics, University of Colorado, 440 UCB, Boulder, CO  80309, USA}
\author{Ana Maria Rey}
\affiliation{JILA, NIST, and Dept. of Physics, University of Colorado, 440 UCB, Boulder, CO  80309, USA}
\affiliation{Center for Theory of Quantum Matter, University of Colorado, Boulder, CO 80309, USA}

\begin{abstract}
We study  the quantum dynamics of many-body  arrays of two-level atoms in a driven cavity subject to collective decay and  interactions mediated by the cavity field. We work in the bad cavity limit accessible, for example, using long-lived electronic clock states of alkaline earth atoms, for which the bare atomic linewidth is much less than the cavity linewidth.  In the absence of interactions, our system reduces to previously studied models of collective fluorescence. We show that  while interactions do not qualitatively change the steady state properties, they lead to a drastic change in the dynamical properties. We find that, for some interval of driving strengths, the system shows two very distinct types of transient behaviors that depend on the initial state of the system. In particular, there is a parameter regime where the  system features oscillatory dynamics with a period of oscillation that becomes much shorter than the duration of the overall transient dynamics as the atom number increases. We use both mean field and exact numerical calculations of the quantum system to investigate the dynamics.
\end{abstract}

\pacs{}
\maketitle

\section{Introduction}

Driven-dissipative systems are ubiquitous in nature and are fundamental to modern quantum science. Even in the classical realm
non-linear driven dissipative systems can display fascinating behavior. For example, a macroscopic number of intrinsically noisy driven coupled oscillators can exhibit the phenomenon of synchronization~\cite{Pikovsky}, which describes the spontaneous locking of the oscillators to a common phase. At  the quantum level their behavior can become extremely complex and intriguing~\cite{Galve}. While dissipation generally tends to degrade quantum correlations, it is now widely appreciated that it can also give rise to many-body physics not possible with strictly coherent dynamics, and can be used explicitly for the creation of entanglement~\cite{Kraus,Diehl}. Harnessing dissipation, and determining its effect on interacting many-body systems is a fundamental step for advancing quantum information and the technological promise of  quantum computers and ultra-precise sensors.

Collective spin models are among the most amenable systems to investigate the interplay between dissipation and interactions, since their reduced Hilbert space allows for exact solutions with current computational resources and at the same time their collective nature increases the parameter regime where simple mean field treatments are valid. Recently, collective spin models and generalizations with  independent decay and dephasing have regained theoretical interest~\cite{DallaTorre, Schneider, Morrison, Kessler,Iemini,Hannukainen,Ferreira,Lang,Maghrebi,Walls,LeeT,Shammah,FossFeig} given impressive  experimental advances in quantum laboratories.  These studies have shown that these systems have steady states with very rich phase diagrams~\cite{Ferreira,Walls},  can undergo dynamical phase transitions~\cite{Morrison,Zunkovic} including self-organized criticality \cite{Helmrich} and bistability~\cite{Gothe, Ritter,Lugiato}, as well as emergent quantum synchronization \cite{Bohnet,Zhu,Meiser}, and for some parameter regimes in the thermodynamic limit,  steady states characterized by oscillating observables~\cite{Carmichael, LeeT,Chan,Buca} in close connection with recently proposed  time-crystals \cite{Iemini,Gong,Tucker}. Spin squeezing has also been a prominent quantity of study in collective systems with strong dissipation~\cite{Liu,Shen,Xu,Wolfe,LewisSwan,Hu}.

In this paper we consider a dissipative version of the Lipkin-Meshkov-Glick~\cite{Lipkin} model that arises in the study of ultra long-lived atoms inside a rapidly decaying QED cavity~\cite{Norcia0}, where the bare atomic linewidth is much less than the cavity linewidth. In this limit the rapidly decaying cavity field can be adiabatically eliminated  giving rise both to effective elastic interactions between the atoms and to a collective decay process responsible, among other things, for superradiance \cite{Norcia}. Cavity mediated interactions alone lead to a steady state where all atoms are de-excited, and as such we incorporate an additional coherent drive to pump energy into the system and thus realize a nontrivial steady state. Such a driven-dissipative system has been studied with  exact and mean field calculations in the case of zero interactions~\cite{Drummond,Drummond2,Drummond3,Carmichael}, where at a critical value of the driving strength a dissipative phase transition was identified in the steady state. Below the critical driving strength there exists a superradiant phase, which features a steady state with non-zero inversion, whilst above this critical value there is a normal phase characterized by zero net inversion and slowly decaying transient oscillations.

Here, we first show (sec.~\ref{SSProperties}) that interactions don't qualitatively change the properties of the steady state and just induce a renormalization of the critical drive and an overall rotation of the spin observables. In line with other reports in the literature~\cite{Schneider,DallaTorre,LeeT}, we also find parameter regimes where the steady state is highly squeezed, scaling with particle number in a similar way to previous studies~\cite{LeeT}.
However, interactions significantly modify the dynamics (sec.~\ref{Dynamics}). Using a mean field formulation we find that there is a window of driving strengths, which disappears  without  interactions, in which the transient dynamics is  strongly  sensitive to  the initial conditions. This results in the existence of three distinct regions, instead of two, as compared to the exact steady state or to the dynamics of the non-interacting system (sec.~\ref{DynamicsA}).

While in general mean field theory describes poorly the long time dynamics, specifically for systems with large enough drive strengths and  finite $N$, we find that the mean field phases remain  present in the exact solution (sec.~\ref{DynamicsB}) by integrating numerically the master equation of the system. Accounting for  quantum correlations we observe  that, in contrast to the mean field predictions, Rabi oscillations do decay (for large enough drives). The decay rate scales much faster with the number of atoms compared to the cases  when there is no dissipation (\cite{Schliemann,Lerose}) or no interactions. In the latter case it reduces to  simple exponential decay. Finally, in sec.~\ref{LOperator} we relate the phase boundaries of the mean field phase diagram with the opening or closing of gaps in the real and imaginary parts of the spectrum of the Liouville operator. In this way, we tie features of the mean field dynamics to properties of the exact quantum solution.

While studies of fluorescence of driven atoms in the presence of dissipation, both collective and independent, have a very long history~\cite{Drummond,Drummond2,Drummond3,Carmichael} and a myriad results about them have been obtained throughout the years, the investigations presented here are timely
since  only very recently, atoms featuring  ultra long-lived optical transitions, such as the clock transition in strontium, have been cooled down and loaded in optical cavities~\cite{Norcia,Norcia0}.  These capabilities are opening  a path for the first experimental observation of dynamical phases in driven dipoles. This is thanks to the extremely long lifetime (larger than a hundred seconds) of the clock transition  which allows the system to attain the predicted  non-trivial steady states before spontaneous decay destroys correlations between atoms. At the same time, the weak dipole moment slows down the collective dynamics  making it to occur  at times  accessible with standard probes in current cavity QED experiments. This is strikingly different to standard dipole allowed transitions which need much faster probes for detection.

\section{Model}\label{Model}
We consider first a system composed of a single cavity mode interacting collectively with $N$ atoms whose dynamics are determined by a master equation:
\begin{equation}\label{IAtomLightMasterEquation}
\frac{d\hat{\rho}}{dt}=-i\bigg[g\Big(\hat{a}\hat{J}^{+}+\hat{a}^{\dagger}\hat{J}^-\Big)-\Delta \hat{a}^{\dagger}\hat{a},\hat{\rho}\bigg]+\kappa\Big(\hat{a}\hat{\rho} \hat{a}^{\dagger}-\frac{1}{2}\{\hat{a}^{\dagger}\hat{a},\hat{\rho}\}\Big),
\end{equation}
where $\{\cdot,\cdot\}$ denotes the anticommutator, $\hat{a}$ is the destruction operator that acts on the cavity mode, $\hat{J}^{\pm}=\hat{J}_x\pm i\hat{J}_y$ and $\hat{J}_{x,y,z}=\frac{1}{2}\sum_i^N\hat{\sigma}_i^{x,y,z}$ are collective pseudospin operators, $\hat{\sigma}_i^{x,y,z}$ are Pauli spin operators that act on the internal two level system of atom $i$, $\Delta$ is the cavity detuning with respect to the atomic transition frequency, $\kappa$ is the cavity decay rate and $g$ is the atom-light coupling constant.

In the bad cavity limit, accessible with long lived dipoles, the spontaneous decay rate of the atoms, $\gamma$, is much smaller than the cavity decay, $\kappa$ and can be neglected, to an excellent approximation. For example in the case of the clock transition in $^{87}$Sr, $\gamma$ is of the order of millihertz while $\kappa$ is typically of the order of kilohertz and thus nine orders of magnitude larger. In the bad cavity limit, $\kappa\gg g$, we can adiabatically eliminate the cavity field from Eq.~(\ref{IAtomLightMasterEquation}) leading to a collective spin model~\cite{Bonifacio},

\begin{equation}\label{IMasterEquation}
\frac{d\hat{\rho}}{dt}=-i\Big[\chi \hat{J}^{+}\hat{J}^-, \hat{\rho}\Big]+\Gamma\bigg(\hat{J}^{-}\hat{\rho} \hat{J}^{+}-\frac{1}{2}\big\{ \hat{J}^{+}\hat{J}^{-},\hat{\rho}\big\}\bigg),
\end{equation}
where
\begin{align}\begin{split}
\chi&=\frac{4g^2\Delta}{4\Delta^2+\kappa^2},\\
\Gamma&=\frac{4g^2\kappa}{4\Delta^2+\kappa^2}.
\end{split}\end{align}
Here, $\chi$ describes effective elastic exchange interactions and $\Gamma$ describes collective coherent decay of the atomic system, responsible for superradiance. As mentioned before, under Eq.~(\ref{IMasterEquation}) the system reaches a steady state with all atoms de-excited so we add a single particle driving term on resonance with the atomic transition and with a Rabi frequency $\Omega$. The full evolution of the system is thus given by
\begin{equation}\label{IMasterEquation2}
\frac{d\hat{\rho}}{dt}=-i\Big[\chi \hat{J}^{+}\hat{J}^-+\Omega\hat{J}_x, \hat{\rho}\Big]+\Gamma\bigg(\hat{J}^{-}\hat{\rho} \hat{J}^{+}-\frac{1}{2}\big\{ \hat{J}^{+}\hat{J}^{-},\hat{\rho}\big\}\bigg)=\mathcal{L}\hat{\rho}.
\end{equation}
The Hamiltonian part, when restricted to the collective manifold, can be reduced to a particular realization of the Lipkin-Meshkov-Glick model:
\begin{equation}\label{IHamiltonian}
\hat{H}=\chi \hat{J}^{+}\hat{J}^-+\Omega \hat{J}_x=\chi \hat{J}^2-\chi \hat{J}_z^2+\chi \hat{J}_z+\Omega \hat{J}_x\approx \chi \frac{N }{2}\Big(\frac{N }{2}+1\Big)-\chi \hat{J}_z^2+\Omega \hat{J}_x,
\end{equation}
where we have neglected $\chi \hat{J}_z$ on account of it being a factor of $N$ smaller than the other terms in $\hat{H}$. Here we focus on initial states that are eigenstates of $\hat{J^2}$, which is  a conserved quantity of the dynamics given by Eq.~(\ref{IMasterEquation2}). Consequently, this term only leads to a uniform energy shift in the Hamiltonian. It is important to point out, however, that the $\hat{J}^2$ term gives some degree of protection against single particle inhomogeneities or slow non-collective noise sources~\cite{Norcia}. Thus, Eq.~(\ref{IHamiltonian}) can also be understood as a one-axis twisting Hamiltonian in the presence of driving. As a final comment, we remark that theoretical studies of the Lipkin-Meshkov-Glick model typically consider $\chi$ and $\Gamma$ to scale as $\frac{1}{N}$ to keep the energy extensive, in contrast to our case in which $\chi$ and $\Gamma$ are independent of $N$. As a consequence, when the number of atoms increases, we observe dynamics that become faster in real time.

\section{Steady State Properties}\label{SSProperties}
\begin{figure}
\includegraphics[width=0.5\textwidth]{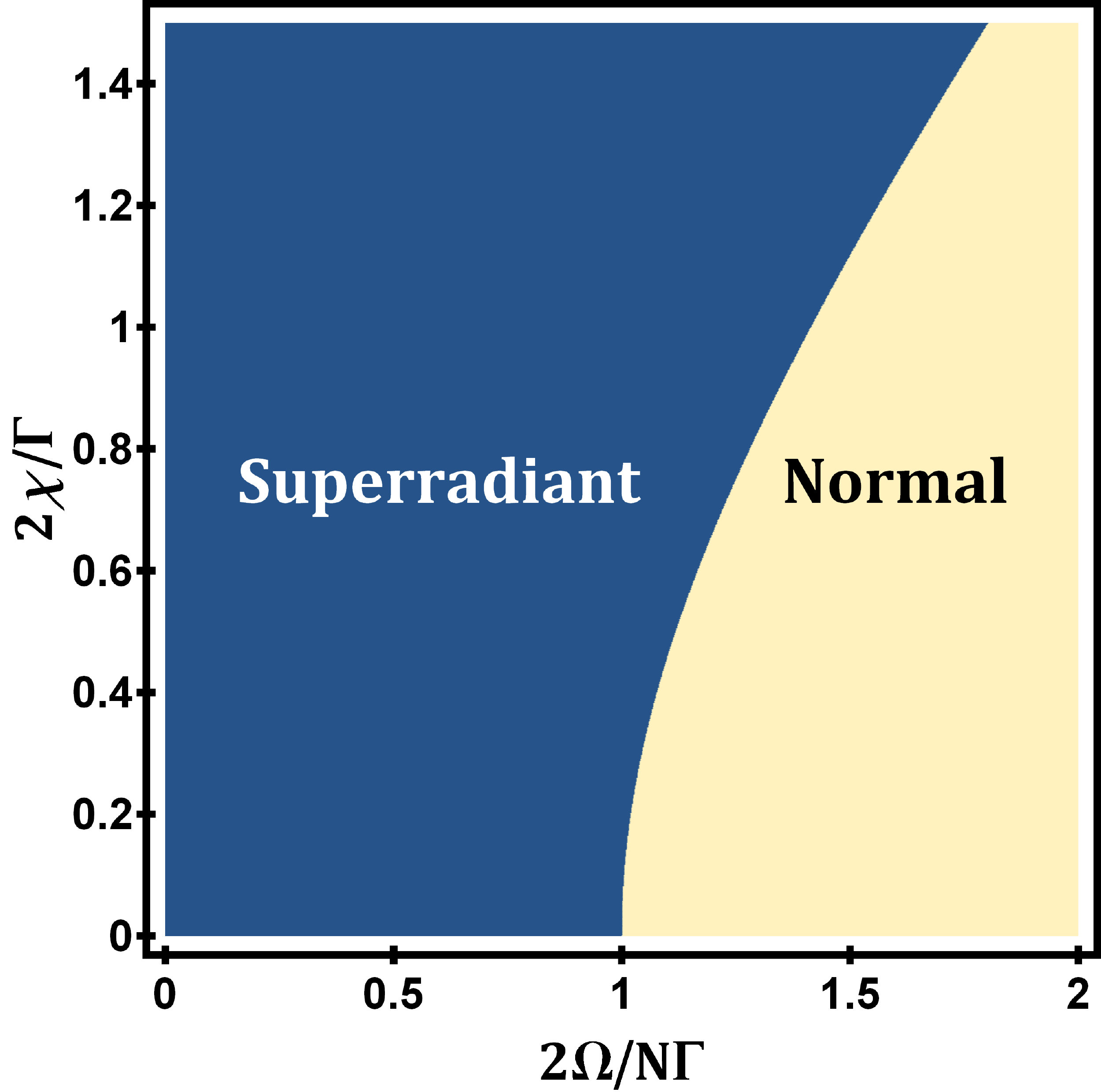}\caption{We show the phase diagram of the system in the $\Omega-\chi$ plane. The transition line is determined by $\frac{2\Omega_c}{N\Gamma}=\sqrt{1+\big(\frac{2\chi}{\Gamma}\big)^2}$. When $\Omega<\Omega_c$, we have the superradiant phase, with nonzero atomic inversion. When $\Omega>\Omega_c$, atomic inversion is 0 and we are in the normal phase.}\label{IIPhaseDiagramSS}
\end{figure}

\begin{figure}
\centering
\includegraphics[width=0.8\textwidth]{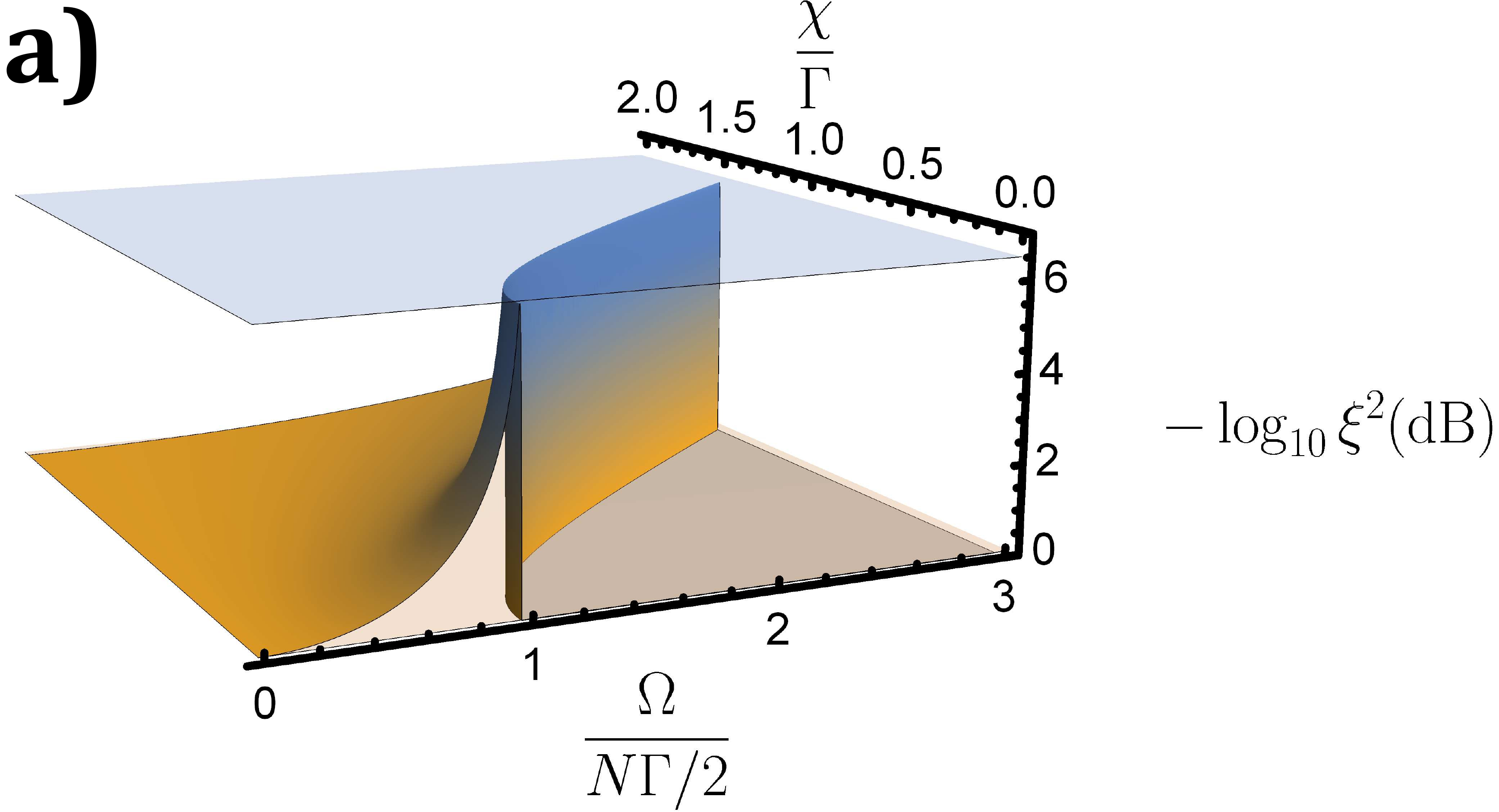}\vspace{0.9cm}
\begin{minipage}{0.48\textwidth}
\includegraphics[width=1\textwidth]{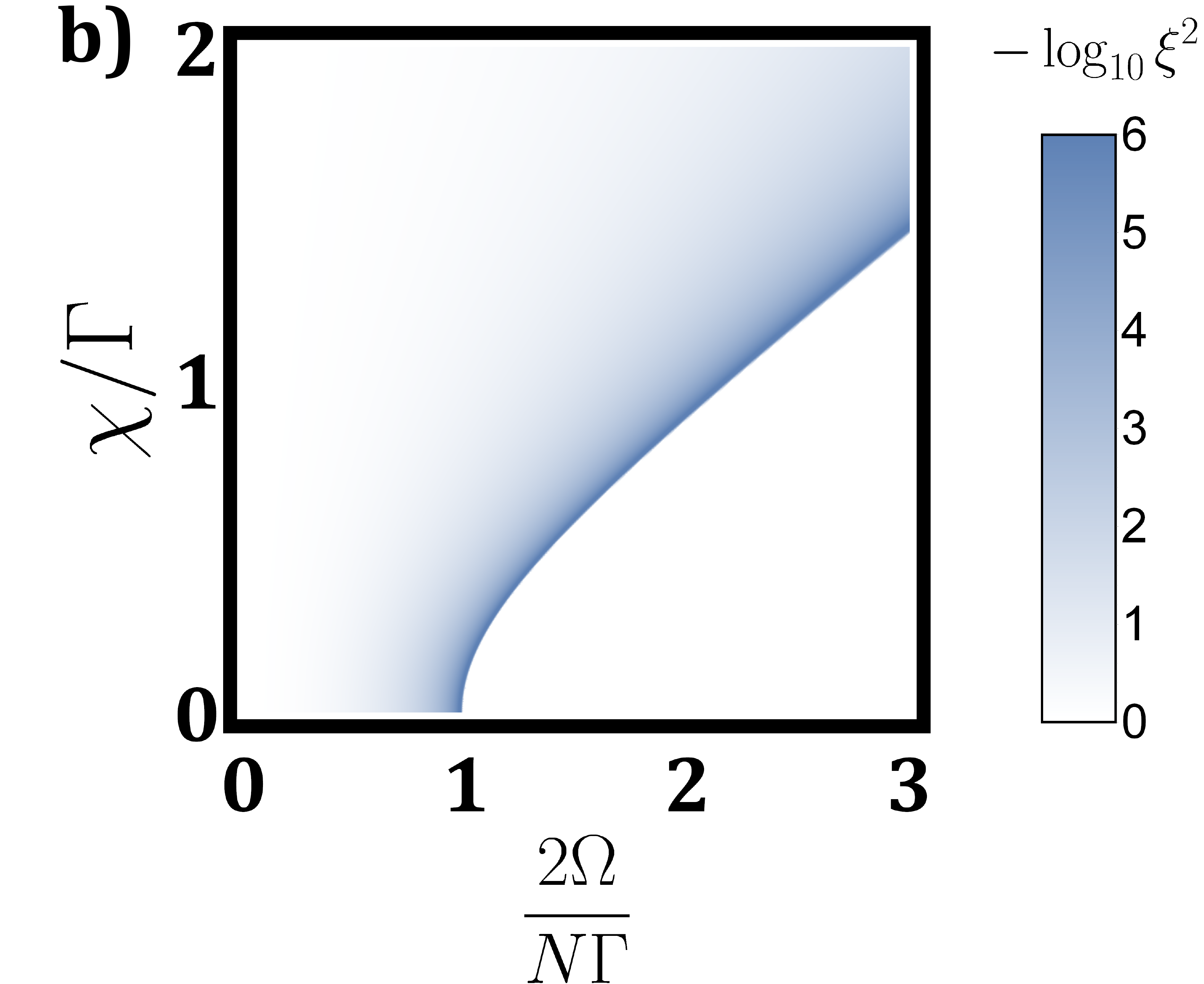}
\end{minipage}\hfill
\begin{minipage}{0.48\textwidth}
\includegraphics[width=1\textwidth]{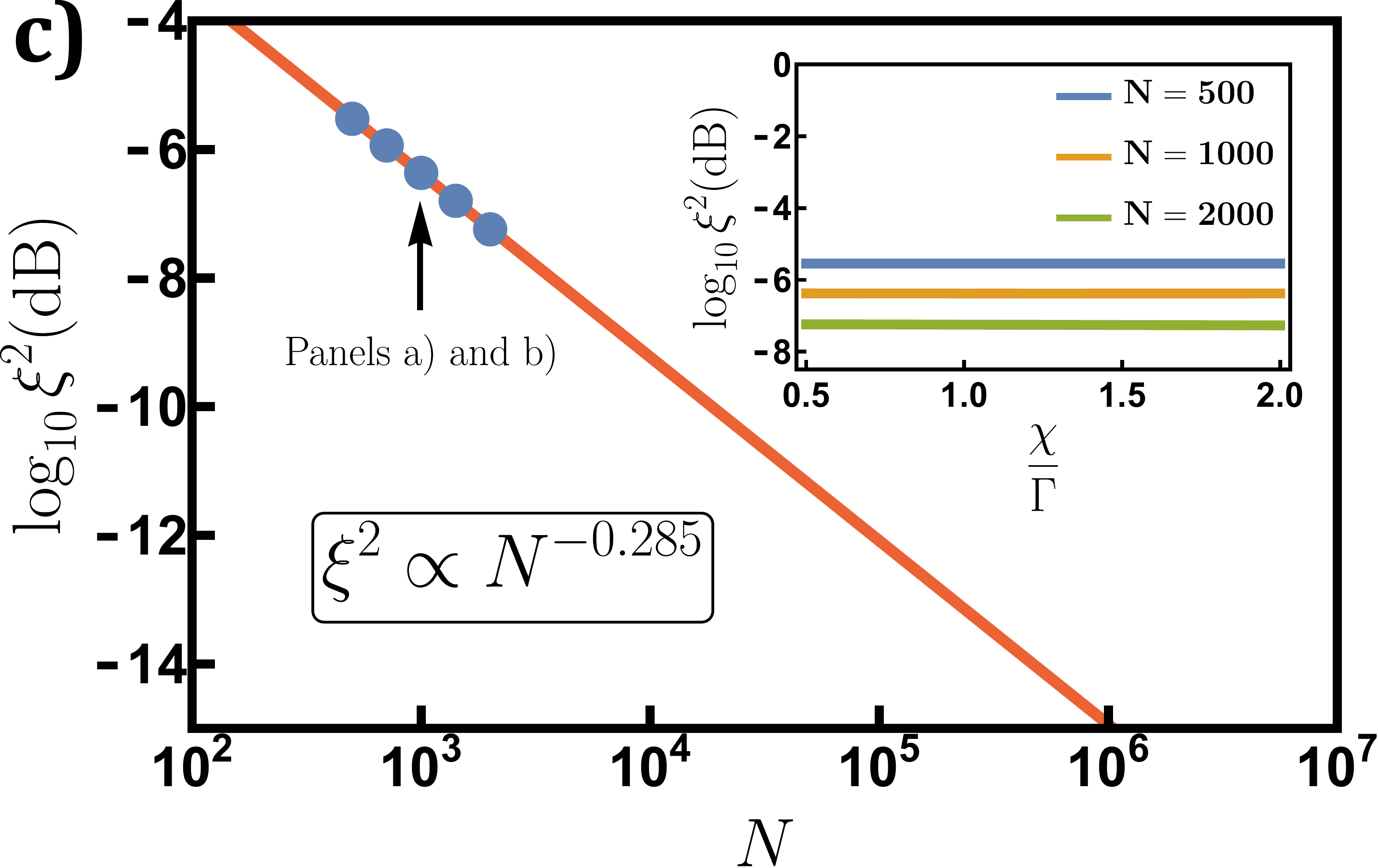}\vspace{0.8cm}
\end{minipage}
\caption{In panel (a), we plot $-\log_{10}\xi^2$ (spin squeezing) as a function of $\frac{2\Omega}{N\Gamma}$ and $\frac{\chi}{\Gamma}$ for $N=1000$. The translucent orange plane marks $\xi^2=1$, while the translucent blue plane indicates the maximum squeezing ($\log_{10}\xi_{\mathrm{max}}^2\approx -6.4$).  At fixed $\frac{\chi}{\Gamma}$, as $\Omega$ is increased from 0, squeezing gets increasingly better reaching a maximum close to $\Omega=\Omega_c$. Beyond $\Omega_c$ squeezing gets worse very quickly. This can be seen also in panel (b) which shows a top down view of panel (a) and further confirms that the best squeezing is obtained for $\Omega$ in a narrow region below $\Omega_c$. Panel (c) shows the scaling of the optimum squeezing as a function of $N$, from which an exponent can be extracted. It is very close to the exponent reported in~\cite{LeeT}. We extrapolate this behaviour because, in modern experiments, attaining $N\approx 10^5$ is feasible~\cite{Norcia0}. For that specific particle numer, squeezing should be at about $12$ dB. The inset shows that the best squeezing is independent of $\chi$, as discussed in the text.} \label{IISteadyStateSqueezing}
\end{figure}

The steady state of the system, $\hat{\rho}_{ss}$ is characterized by $\mathcal{L}\hat{\rho}_{ss}=0$.
To solve it, we use the methods of~\cite{Drummond3,Puri} and introduce displaced spin operators: $\tilde{J}^+=\hat{J}^+-\alpha \hat{I}$, where $\hat{I}$ is the identity matrix. If we choose:
\begin{equation}
\alpha=\frac{i\Omega}{\Gamma-2i\chi},
\end{equation}
then the master equation simplifies to:
\begin{equation}
-i\chi\bigg(\tilde{J}^{+}\tilde{J}^{-}\hat{\rho}_{ss}-\hat{\rho}_{ss}\tilde{J}^{+}\tilde{J}^{-}\bigg)+\Gamma\bigg(\tilde{J}^{-}\hat{\rho}_{ss} \tilde{J}^{+}-\frac{1}{2} \tilde{J}^{+}\tilde{J}^{-}\hat{\rho}_{ss}-\frac{1}{2}\hat{\rho}_{ss} \tilde{J}^{+}\tilde{J}^{-}\bigg)=0.
\end{equation}
From it, it can be seen that the ansatz:
\begin{equation}
\hat{\rho}_{ss}\propto\frac{1}{\tilde{J}^-}\frac{1}{\tilde{J}^+}
\end{equation}
solves independently the Hamiltonian and dissipative parts. Thus, a closed form expression can be written for the system's density matrix:
\begin{equation}\label{IIrho}
\hat{\rho}_{ss}=\mathcal{C}|\alpha|^2\bigg(\frac{1}{\hat{J}^{-}-\alpha^* \hat{I}}\bigg)\bigg(\frac{1}{\hat{J}^{+}-\alpha \hat{I}}\bigg),
\end{equation}
where $\mathcal{C}$ is a normalization factor that ensures $\Tr(\hat{\rho})=1$. This allows us to calculate, for example, the expectation value of $\hat{J}_z$ in the large $N$ limit (keeping $\frac{\Omega}{N}$ fixed). The result is:
\begin{equation}\label{IIOrderParameter}
\braket{\hat{J}_z}_{ss} =
\begin{cases}
  -\frac{N}{2}\sqrt{1-\frac{\Omega^2}{\Omega_c^2}}&\Omega<\Omega_c\\[0.4cm]
 ~ ~ ~~ 0 &\Omega>\Omega_c,
\end{cases}
\end{equation}
where
\begin{equation}
\Omega_c=\frac{N}{2}\sqrt{\Gamma^2+4\chi^2}.
\end{equation}
Equation (\ref{IIOrderParameter}) shows that there is a phase transition at $\Omega=\Omega_c$ for which $\braket{\hat{J}_z}_{ss}$ is a good order parameter. We refer to these phases as superradiant $({\mathcal S})$ and normal $({\mathcal N})$ for $\Omega<\Omega_c$ and $\Omega>\Omega_c$ respectively (see Fig.~\ref{IIPhaseDiagramSS}), consistent with previous studies in the absence of elastic interactions~\cite{Carmichael}.
These works also found phenomena such as spin squeezing~\cite{Kitagawa}, which we find persists in the presence of elastic interaction as shown in Fig.~\ref{IISteadyStateSqueezing}. There, we use the definition of the squeezing parameter $\xi $ introduced by Wineland et al. \cite{Wineland}:
\begin{equation}\label{SqueezingDef}
\xi^2=\underset{\mathbf{n_{\perp}}}{\mathrm{min}}\bigg(\frac{N\braket{\Delta \hat{J}_{\mathbf{n_\perp}}^2}}{|\braket{\mathbf{\hat{J}}}|^2}\bigg),
\end{equation}
$\braket{\mathbf{\hat{J}}}=\big(\braket{\hat{J}_x},\braket{\hat{J}_y},\braket{\hat{J}_z}\big)$ is the expectation value of the spin vector, $\braket{\Delta \hat{J}^2_n}=\braket{(\mathbf{\hat{J}}\cdot\mathbf{n})^2}-(\braket{\mathbf{\hat{J}}\cdot\mathbf{n}})^2$ is the variance of the projection of the spin in direction $\mathbf{n}$, $\mathbf{n_{\perp}}$ is a direction orthogonal to $\braket{\mathbf{\hat{J}}}$ (that is, $\mathbf{n_{\perp}}\cdot\braket{\mathbf{\hat{J}}}=0$) and we are minimizing the quantity in parentheses in Eq.~(\ref{SqueezingDef}) among all possible $\mathbf{n_{\perp}}$ directions. If $\xi^2<1$, then the system is entangled~\cite{Sorensen}.
In Fig.~\ref{IISteadyStateSqueezing} we plot $-\log_{10}\xi^2$ and it clearly shows that the phases defined by Eq.~(\ref{IIOrderParameter}) have very different squeezing properties: it is positive in the ${\mathcal S}$ phase (thus entangled) and negative in the ${\mathcal N}$ phase.
The best squeezing is obtained in a very small region around the transition line and is, quite surprisingly, independent of the interaction strength $\chi$. This should become clear by looking at Eq.~(\ref{IIrho}), from which we can deduce that the squeezing can only be a function of $|\alpha|$ so that any change in $\chi$ can be compensated by a change in $\Omega$ that keeps $|\alpha|$ constant. 
\\

\begin{figure}

\subfloat[]{\includegraphics[width=0.3\textwidth,trim={2.1cm 2.5cm 3.5cm 2.5cm},clip]{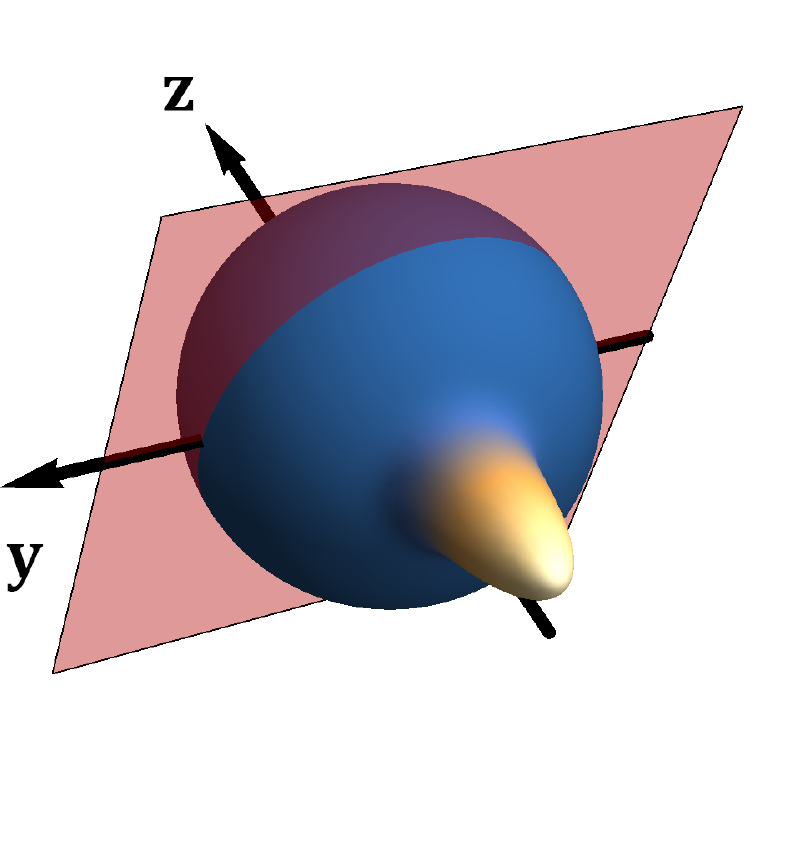}}
\subfloat[]{\includegraphics[width=0.3\textwidth,trim={2.1cm 2.5cm 3.5cm 2.5cm},clip]{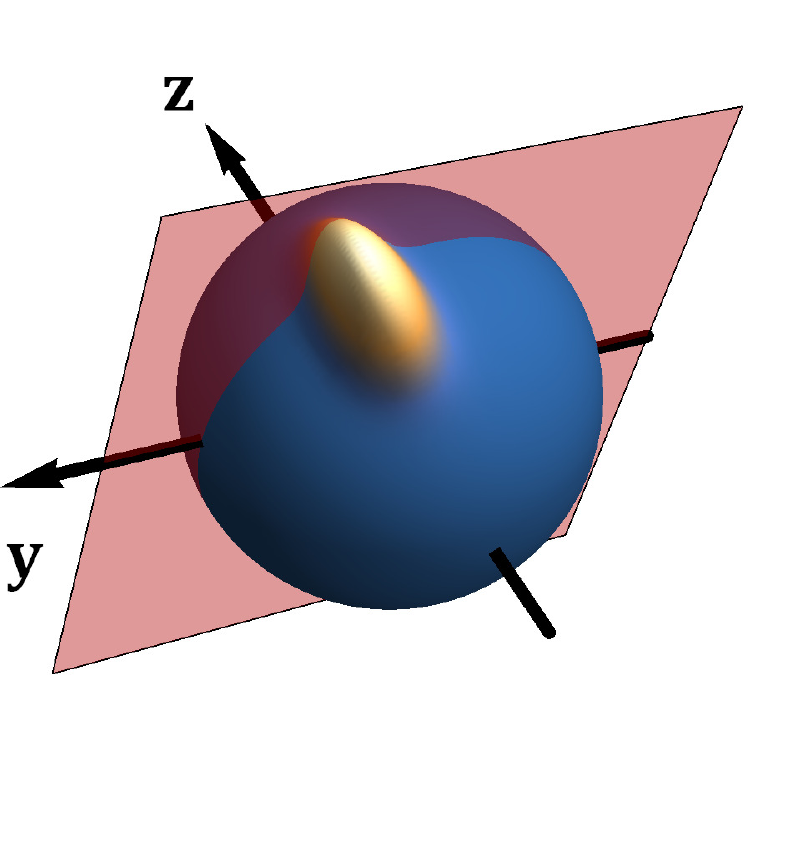}}
\subfloat[]{\includegraphics[width=0.3\textwidth,trim={2.1cm 2.5cm 3.5cm 2.5cm},clip]{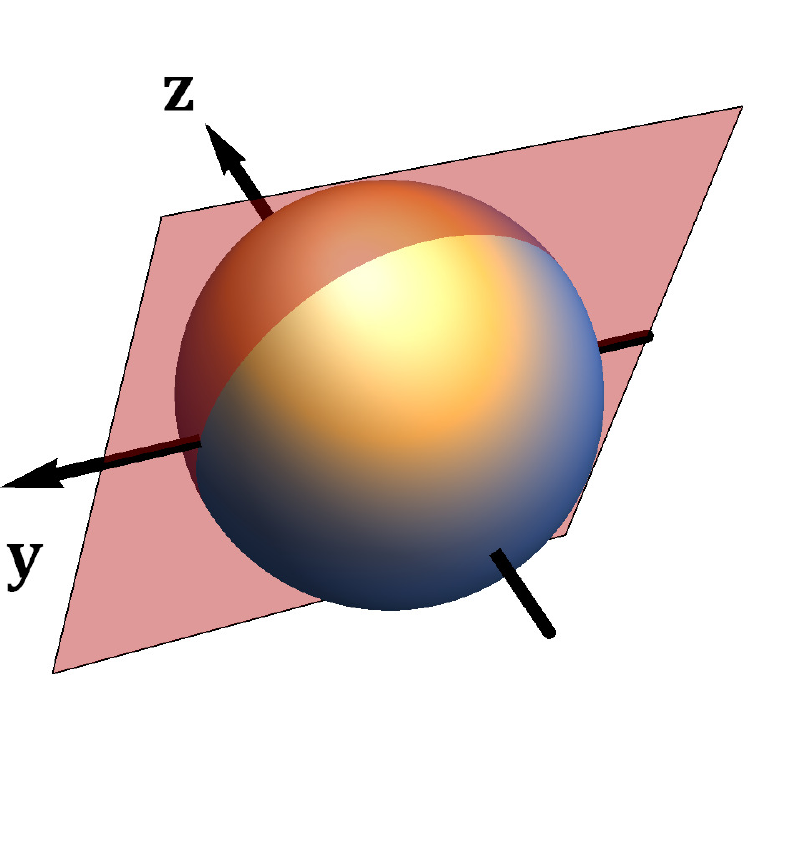}}
\caption{Husimi distributions of the steady state for different values of $\Omega$. a) $\Omega=0.45\Omega_c$, b) $\Omega=0.96\Omega_c$, c) $\Omega=1.79\Omega_c$. The size of the bulge protruding from the sphere is proportional to the length of the Bloch vector and the translucent plane is the XY plane.} \label{IISqueezingSpheres}
\end{figure}

To get more insight into the squeezing we calculate the Husimi distribution, defined by
\begin{equation}
Q(\theta,\phi)=\frac{(N+1)}{4\pi}\bra{\theta,\phi}\hat{\rho}\ket{\theta,\phi},
\end{equation}
where $\ket{\theta,\phi}$ is a spin coherent state, defined as the eigenstate of $\mathbf{n}\cdot\mathbf{\hat{J}}$ with positive eigenvalue, where $\mathbf{n}=(\sin\theta\cos\phi,\sin\theta\sin\phi,\cos\theta)$. We show the Husimi distributions of states in different regions of the phase diagram in Fig.~\ref{IISqueezingSpheres}.
Inside the $\mathcal{S}$ region, the state is very close to being a spin coherent state, with the noise ellipse close to being isotropic. When $\Omega=0$, the steady state is clearly a coherent state pointing in the $-\hat{z}$ direction, so the inclusion of a small driving just lifts this state a little bit but doesn't change its qualitative properties. The steady state is reached in a situation where the rate at which the state is lifted by $\Omega$ is compensated by the rate at which the state relaxes back to $-\hat{z}$ due to dissipation. If there were no interactions, the state would be pointing in the $yz$ plane (perpendicular to the driving axis), but $\chi$ induces a rotation about the $\hat{z}$ axis. As we increase $\Omega$ and move closer to the transition region, the driving starts winning over the dissipation shearing the noise ellipse in the vertical direction and moving the collective Bloch vector closer to the equator since a larger dissipation is needed to equilibrate the system. Note that dissipation is the greatest at the equator. Finally, when increasing $\Omega$ past the transition point, dissipation is not strong enough to equilibrate the drive at any point in the sphere and the Bloch vector keeps rotating. Dissipation then just decoheres the state, which results in the state spreading over a large area of the Bloch sphere.

\section{Dynamics}\label{Dynamics}
Given that a simple closed form solution of the exact many-body dynamics is not possible, we instead use a mean field analysis and then compare it to the exact steady state presented in  the previous section and the results from exact numerical solutions. The exact equations of motion for the expectation values of the spin observables derived from Eq.~(\ref{IMasterEquation}) are: 
\begin{align}\begin{split}\label{IIIExactDynamics}
\frac{d\braket{\hat{J}_x}}{dt}&=\chi\braket{\{\hat{J}_y,\hat{J}_z\}}+\frac{\Gamma}{2}\braket{\{\hat{J}_x,\hat{J}_z\}}-\frac{\Gamma}{2}\braket{\hat{J}_x},\\
\frac{d\braket{\hat{J}_y}}{dt}&=-\Omega\braket{\hat{J}_z}-\chi\braket{\{\hat{J}_x,\hat{J}_z\}}+\frac{\Gamma}{2}\braket{\{\hat{J}_y,\hat{J}_z\}}-\frac{\Gamma}{2}\braket{\hat{J}_y},\\
\frac{d\braket{\hat{J}_z}}{dt}&=\Omega \braket{\hat{J}_y}-\Gamma\braket{\hat{J}_x^2+\hat{J}_y^2}-\Gamma \braket{\hat{J}_z}.\\
\end{split}\end{align}
\subsection{Mean field analysis}\label{DynamicsA}
The mean field approximation assumes the spin variables are uncorrelated and replaces $\braket{\hat{J}_i\hat{J}_k}\approx\braket{\hat{J}_i}\braket{\hat{J}_k}$ in Eq.~(\ref{IIIExactDynamics}). Furthermore, for convenience we define $s_i=\frac{\braket {\hat{J}_i}}{N/2}$. At the mean field level the equations for $s_i$ are then:
\begin{align}\begin{split}\label{IIIMeanFieldEquations}
\dot{s}_x&=N\chi s_y s_z+\frac{N\Gamma}{2}s_xs_z,\\
\dot{s}_y&=-\Omega s_z-N\chi s_x s_z +\frac{N\Gamma}{2}s_y s_z,\\
\dot{s}_z&=\Omega s_y-\frac{N\Gamma}{2}(s_x^2+s_y^2),
\end{split}\end{align}
and we have assumed here that $N\Gamma$, $N\chi$ and $\Omega$ are comparable. Regardless of the  nonlinear character of the mean field equations, the solution  can be parameterized in terms of a variable $b(t)$:
\begin{align}\begin{split}\label{IIParameterization}
s_{z}(t)&=-\dot{b}(t),\\
s_{+}(t)&=s_{x}(t)+i s_{y}(t)=s_{+}(0)e^{-\big(\frac{N\Gamma}{2}-iN\chi\big)b(t)}+\frac{2i\Omega}{N\Gamma-2iN\chi}\bigg(1-e^{-\big(\frac{N\Gamma}{2}-iN\chi\big)b(t)}\bigg).
\end{split}\end{align}
The first line of Eq.~\ref{IIParameterization} defines $b(t)$, which leaves us free to choose its initial value. The second line of Eq.~\ref{IIParameterization} is obtained through an integration of the equation for $s_+(t)$ and setting $b(0)=0$. Thus, $b(t)$ satisfies:
\begin{align}\begin{split}
b(0)&=0,\\
\dot{b}(t)^2&+V(b(t))=1.
\end{split}\end{align}
where
\begin{align}\begin{split}
 V(b)=\frac{4\frac{\Omega^2}{N^2}}{\Gamma^2+4\chi^2}\bigg(1&-2e^{-\frac{N\Gamma}{2}b}\cos(N\chi b)+e^{-N\Gamma b}\bigg)+|s_+(0)|^2e^{-N\Gamma b(t)}\\[8pt]
&+2|s_+(0)|\sqrt{\frac{4\frac{\Omega^2}{N^2}}{\Gamma^2+4\chi^2}}e^{-\frac{N\Gamma b}{2}}\bigg(\cos(\phi-N\chi b)+e^{-\frac{N\Gamma b}{2}}\cos(\phi)\bigg),\end{split}\end{align}
and
\begin{equation}
\phi=\arg\bigg(\frac{2i\Omega}{N\Gamma-2iN\chi}s_+^*(0)\bigg).
\end{equation}
The time evolution of $b(t)$ can then be understood as the coordinate of a particle of unit mass and energy moving in the effective potential $V(b)$, which clearly depends on the initial conditions. This analogy makes it easier to visualize the solution.

\begin{figure}
\subfloat{\includegraphics[width=0.48\textwidth]{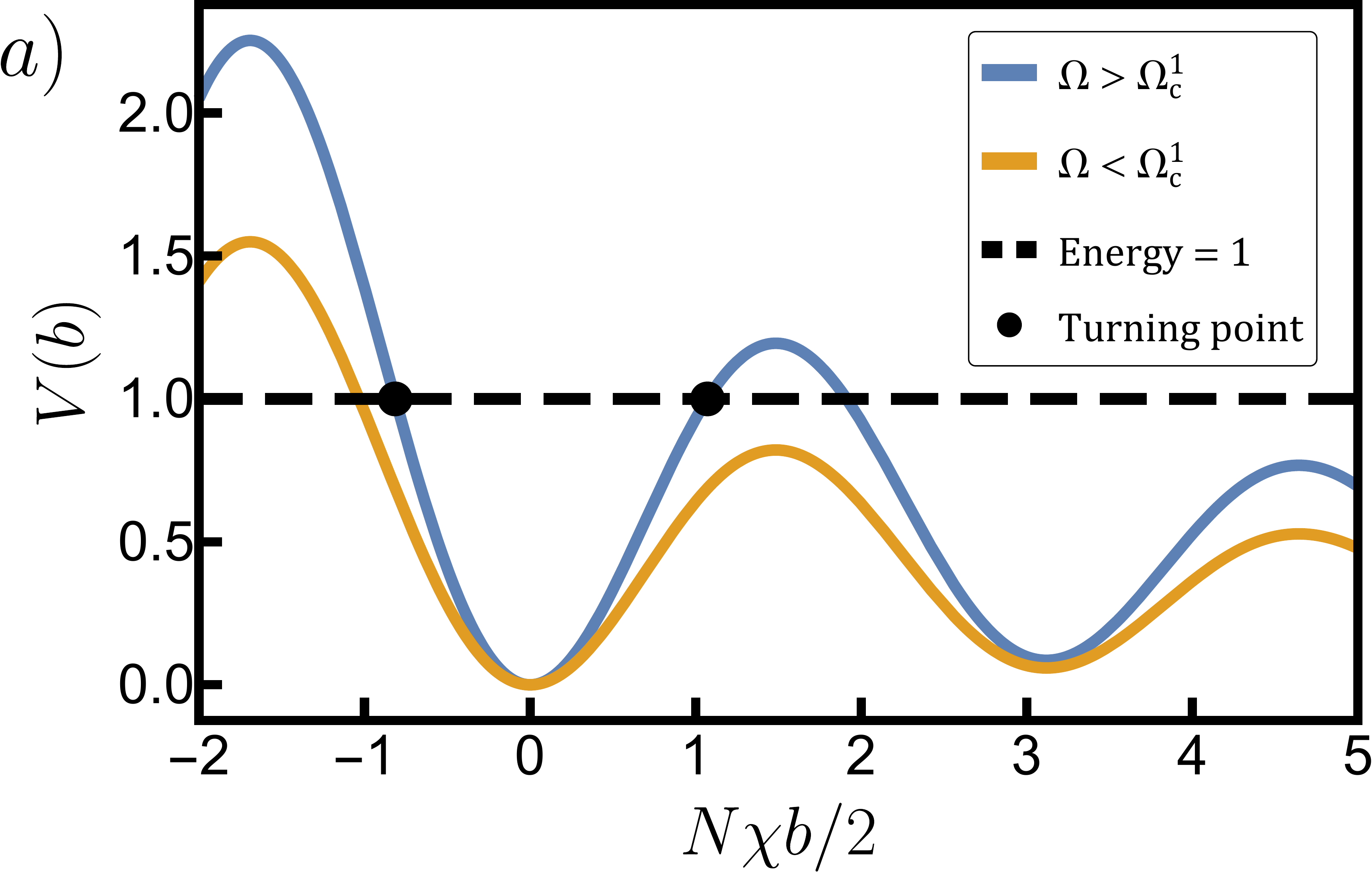}\hspace{0.6cm}}
\subfloat{\includegraphics[width=0.48\textwidth]{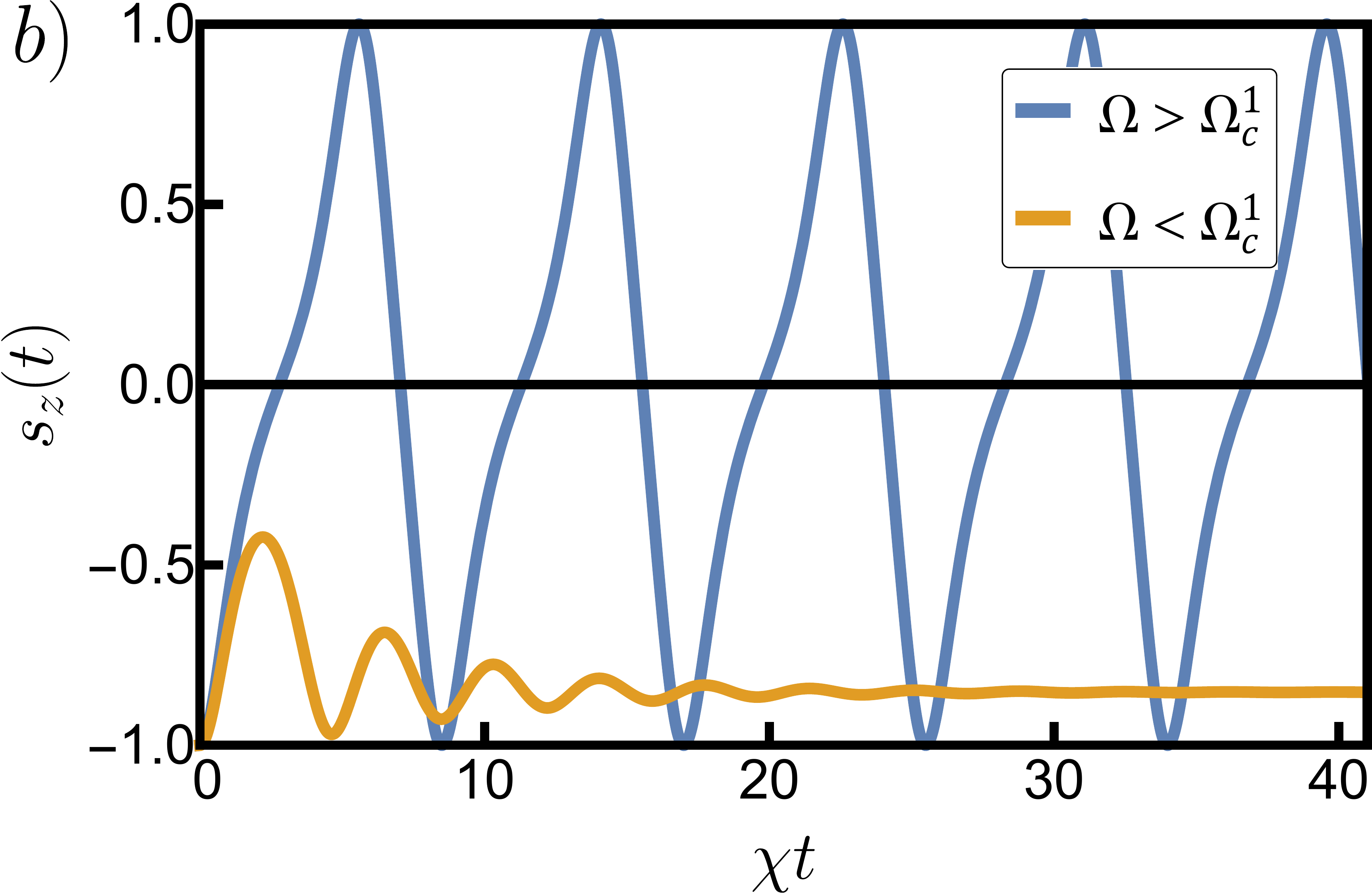}}
\caption{Panel (a) shows the effective potential for $\Omega<\Omega^1_c$ (orange) and $\Omega>\Omega^1_c$ (blue). $b$ is a variable in terms of which all spin projections can be expressed (see Eq.~\ref{IIParameterization}). It starts at 0 and with $\dot{b}>0$ (since $s_z(0)<0$). In the former case, $b$ is not constrained by the potential, can increase without bound and $\dot{b}$ reaches a constant value at long times. In the latter, $b$ is constrained to remain between the turning points shown in the figure, thus showing oscillatory behaviour.  Panel (b) shows the time evolution of $s_z(t)$ for $\Omega<\Omega^1_c$ (orange, decaying) and $\Omega>\Omega^1_c$ (blue, oscillatory).} \label{IIIEffectivePotential}
\end{figure}

In the case of starting with all the atoms de-excited (that is, the south pole), $s_+(0)=0$ and the equations simplify to
\begin{align}\begin{split}
b(0)&=0,\\
\dot{b}(t)^2&+\frac{4\frac{\Omega^2}{N^2}}{\Gamma^2+4\chi^2}\bigg(1-2e^{-\frac{N\Gamma}{2}b(t)}\cos(N\chi b(t))+e^{-N\Gamma b(t)}\bigg)=1.
\end{split}\end{align} As is shown in Fig.~\ref{IIIEffectivePotential}, there are two different situations depending on whether $\Omega$ is greater or smaller than some $\Omega_c^1$ (not equal to the  critical value $\Omega_c$ derived before for   the steady state).  For $\Omega<\Omega_c^1$: $b$ is unbounded, so $\dot{b}$ never changes sign. Thus $s_z$ remains negative throughout the evolution and tends to a constant value in the infinite future which agrees with the exact value calculated from the steady state density matrix. For $\Omega>\Omega_c^1$: $b$ keeps oscillating indefinitely and so does $s_z$. There is no equilibration and the system never reaches a steady state.

\begin{figure}
\includegraphics[width=1\textwidth]{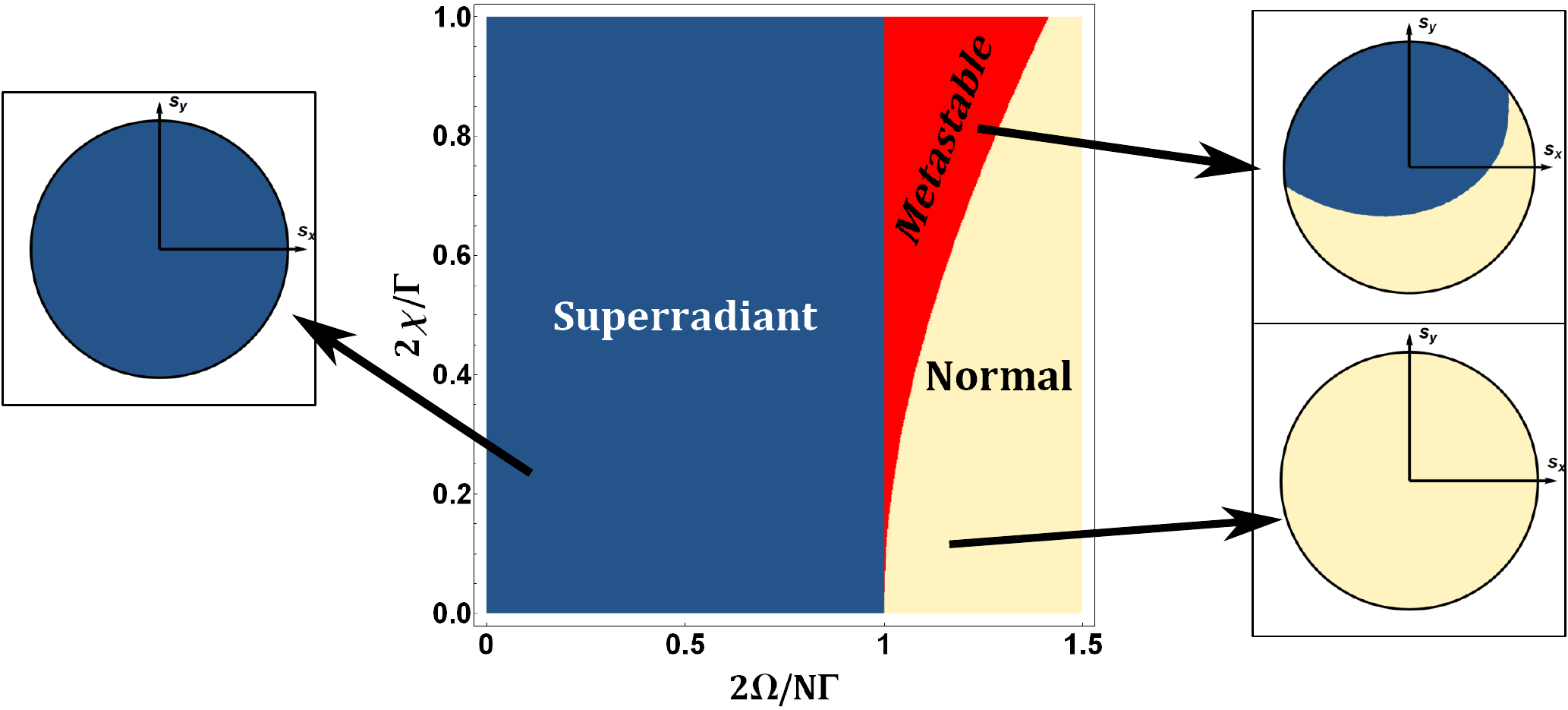}
\caption{Phase diagram constructed from the mean field dynamics. Any point inside a circle corresponds to a specific initial condition of the Bloch vector. Blue indicates the initial conditions which lead to an  inversion that decays to a final steady state  and yellow the initial conditions that lead to an oscillatory behavior. This divides the $\Omega-\chi$ plane in three regions: The blue region featuring  only decaying solutions (circle is fully blue) which lies  inside the superradiant $\mathcal{S}$ steady state region. The red region which features  both decaying and oscillatory solutions (circle is partially blue and partially yellow). We refer to it as the  multistable  regime because of the existence of many different (non-unstable) steady states in the mean field limit. It still lies  inside the superradiant $\mathcal{S}$ steady state region. The yellow region which only features  oscillatory solutions (purely yellow circle) lies  inside the normal  $\mathcal{N}$ steady state region.}\label{IIPhaseDiagram}
\end{figure}

The importance of $\Omega_c^1$ is not its precise value but instead two facts: (1) it depends on the initial conditions (since $V(b)$ itself depends on them), and (2) as all different initial conditions are explored, $\Omega_c^1$ acquires values in the finite interval $\big[N\Gamma/2,\Omega_c\big]$. As a result, when $\Omega<N\Gamma/2$, the system relaxes to the steady state for any initial condition. When $N\Gamma/2<\Omega<\Omega_c$, whether the system relaxes or oscillates indefinitely depends on the initial condition chosen, and when $\Omega_c<\Omega$, the system oscillates indefinitely no matter the initial condition (see Fig.~\ref{IIPhaseDiagram}).
To justify this, we computed the possible steady states predicted by Eqs.~(\ref{IIIMeanFieldEquations}), analyzed the behaviour of small perturbations and found that:

1) For $\Omega<\frac{N\Gamma}{2}$, there is one set of steady states with spin observables given by
\begin{align}\begin{split}
s_+^{ss}&=\frac{2i\Omega}{N\Gamma-2iN\chi},\\[0.2cm]
s_z^{ss}&=\pm\sqrt{1-\frac{4\Omega^2}{N^2\Gamma^2+4N^2\chi^2}}=\pm\sqrt{1-\frac{\Omega^2}{\Omega_c^2}},
\end{split}\end{align}
and stability eigenvalues:
\begin{equation}\label{IIIDecayEigenvalue}
\lambda=N\bigg(\frac{\Gamma}{2}\pm i\chi\bigg)s_z^{ss}.
\end{equation}
Thus, only the state with $s_z^{ss}<0$ is stable and we denote it by $SS_1$.

2) For $\frac{N\Gamma}{2}<\Omega<\Omega_c$, $SS_1$ is still a valid steady state but there is another set of solutions, with spin observables given by
\begin{align}\begin{split}
s_z^{ss}&=0,\\[0.2cm]
s_y^{ss}&=\frac{N\Gamma}{2\Omega},\\[0.2cm]
s_x^{ss}&=\pm\sqrt{1-\frac{N^2\Gamma^2}{4\Omega^2}},
\end{split}\end{align}
and stability eigenvalues:
\begin{equation}
\lambda^2=-\Omega^2\Bigg(\sqrt{1-\frac{N^2\Gamma^2}{4\Omega^2}}+\sgn(s_x)\frac{N\chi}{\Omega}\Bigg)\sqrt{1-\frac{N^2\Gamma^2}{4\Omega^2}}.
\end{equation}
One of the solutions, which we call $SS_2$, is unstable in the region $\frac{N\Gamma}{2}<\Omega<\Omega_c$ while the other one ($SS_3$) has imaginary eigenvalues.  Thus, if the system starts close to $SS_3$, then it will keep oscillating indefinitely (as found numerically), and if it starts close to $SS_1$ it will decay towards it.

3) For $\Omega_c<\Omega$, $SS_1$ is no longer a valid steady state, so only $SS_2$ and $SS_3$ remain. In this case, however, both have imaginary stability eigenvalues, so the system will oscillate, no matter where it starts.

This allows the construction of the phase diagram shown in Fig.~\ref{IIPhaseDiagram}. For $\Omega<N\Gamma/2$ (blue) we have the superradiant region, where the system always decays to a steady state with a time constant $t_{\mathrm{decay}}\sim N^{-1}$ (see Eq.(\ref{IIIDecayEigenvalue})), hence the name superradiant. For $N\Gamma/2<\Omega<\Omega_c$ (red), we have a region of multistability, with more than one possible steady state. Finally for $\Omega>\Omega_c$ (yellow), we have a region where the driving term dominates and generates  oscillations in the spin inversion that can be understood from a single particle perspective, hence the name normal. Note that the boundary between the red and the yellow regions is the same as the boundary  between the superradiant ${\mathcal S}$ and the normal   ${\mathcal N}$  steady state phases, with the yellow  phase exactly corresponding to the ${\mathcal N}$ phase.

\begin{figure}
\includegraphics[width=0.5\textwidth]{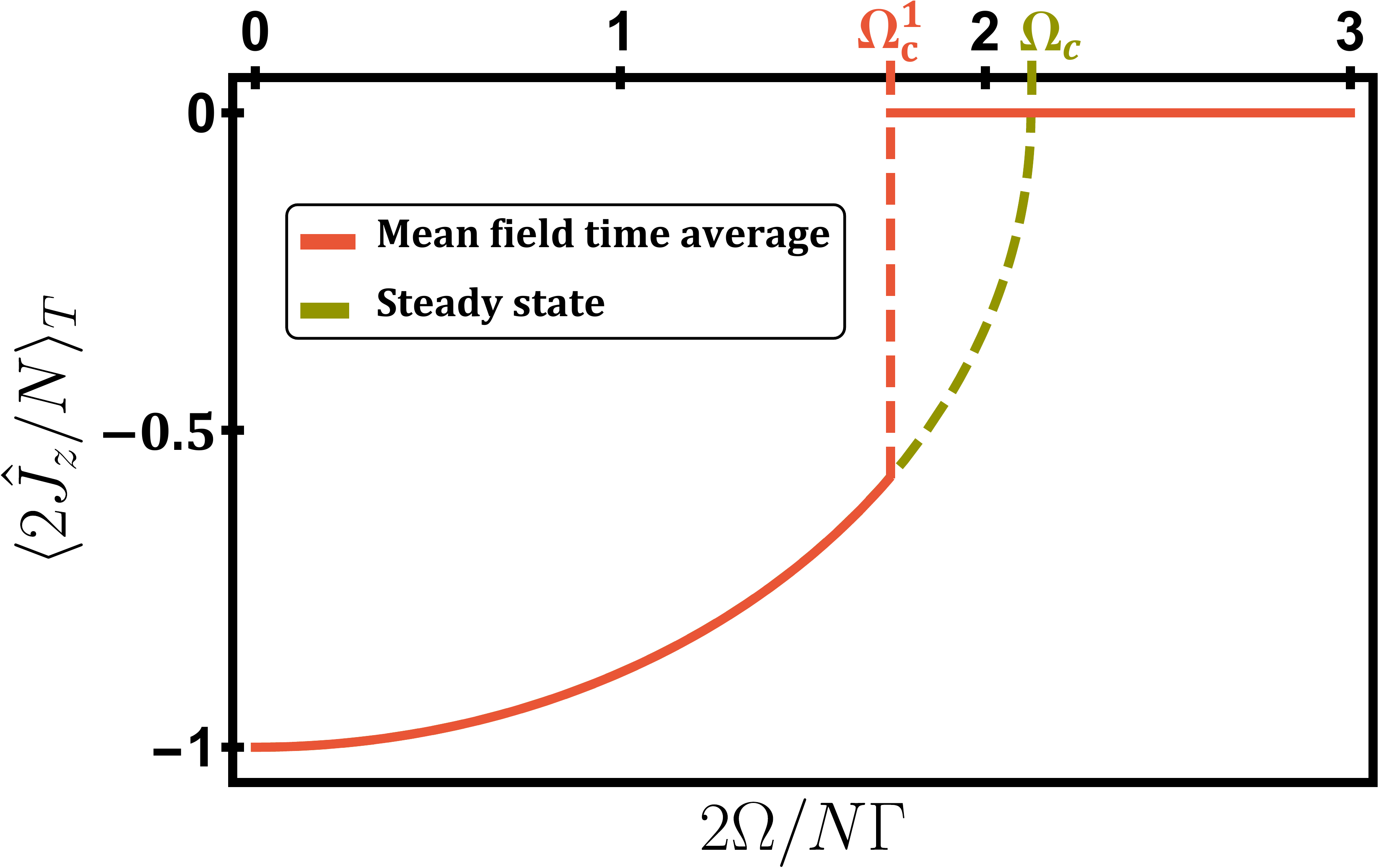}\caption{When $\Omega< \Omega_c^1$, the long time average is equal to the (exact) steady state value. Beyond $\Omega_c^1$, oscillations average  the inversion to 0.}\label{IIFirstOrderTransition}
\end{figure}

Furthermore, if we do a time average of $s_z$,
\begin{equation}
\bar{s}_z=\underset{T\rightarrow\infty}{\lim} \frac{1}{T}\int_0^Ts_z(t)dt,
\end{equation}
 as is usually done  in the context dynamical phase transitions~\cite{Zunkovic}, we find a jump at $\Omega_c^1$ (shown in Fig.~\ref{IIFirstOrderTransition}), characteristic of a first order transition. This can be understood as follows: in the decaying region, the average is always going to be dominated by the steady state value of $s_z$, so both the average and steady state calculations agree. This does not hold all the way to $\Omega_c$, however. Instead, once the oscillating behaviour sets in, $s_z$ will average to 0.

The mean field predictions   and in particular the oscillatory dynamics in the normal  phase seem to contradict the steady state exact solution. In fact, while mean field works well for $\Omega<\frac{N\Gamma}{2}$ and accurately predicts the steady state values of the spin observables, it  completely misses the relaxation process for $\Omega>\Omega_c$. In between these two values, depending on the initial condition, the mean field treatment may or may not describe the relaxation towards the steady state.
\subsection{Exact solution}\label{DynamicsB}
To investigate the extent to which the mean field behaviour is correct, we numerically integrate the master equation for $N=400$. We fix $\chi$ and $\Gamma$, vary $\Omega$ and start at the south pole of the Bloch sphere.
The resulting plots of inversion, shown in Fig.~\ref{IIIInversion}, reveal many interesting features:
\begin{figure}
\includegraphics[width=0.98\textwidth]{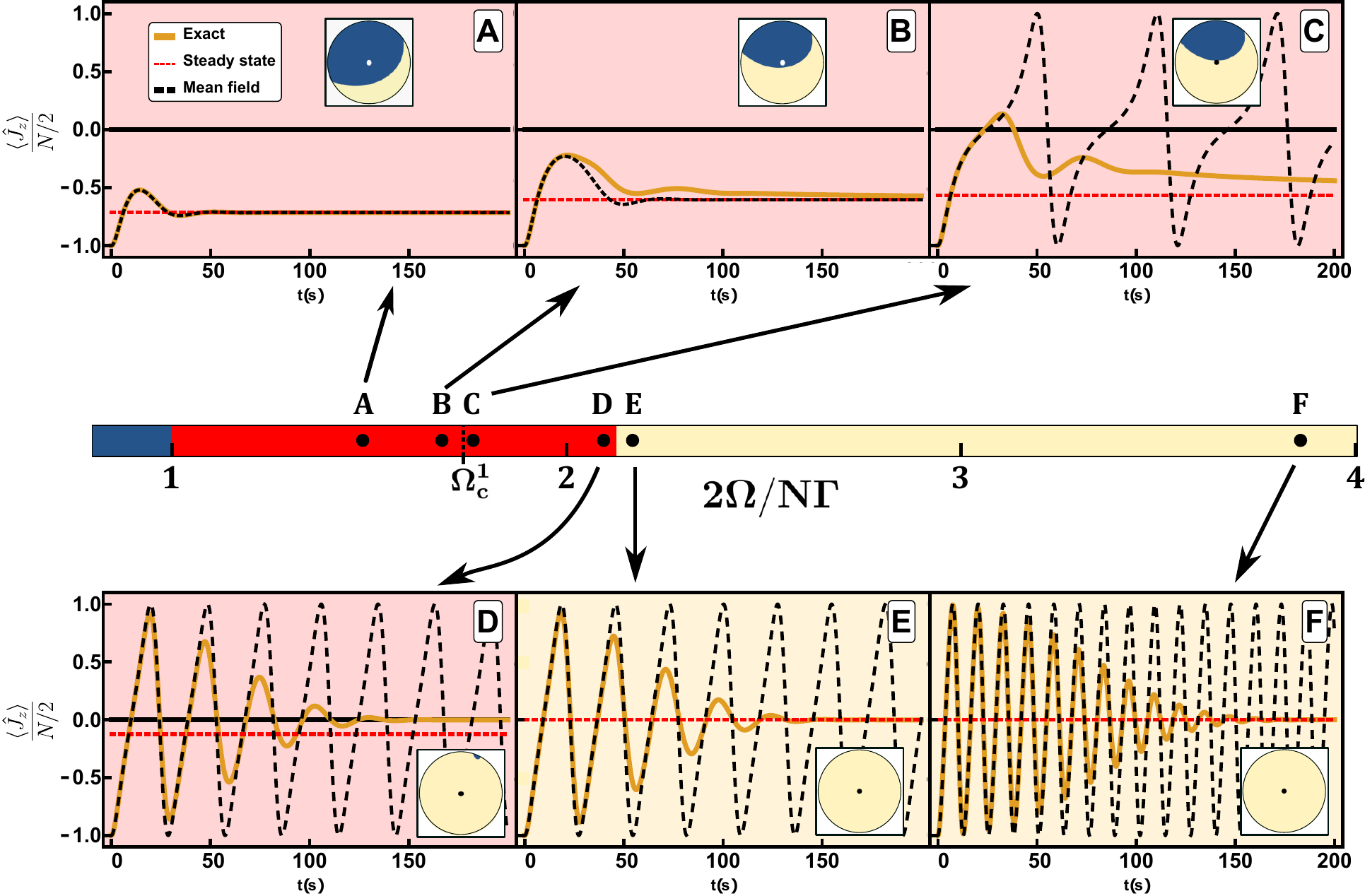}
\caption{The main panels show the results of the exact simulations for $N=400$, $2\chi=1.875\Gamma$, $\Gamma^{-1}=1485$ s and six values of $\frac{2\Omega}{N\Gamma}$: $1.48$ (A), $1.69$ (B), $1.76$ (C), $2.09$ (D), $2.17$ (E) and $3.86$ (F). The insets show  the initial condition (dot in the center), the region of initial conditions in which the mean field solution decays (blue) and in which it oscillates (yellow). The critical values of $\Omega$ for the plotted initial condition are $\frac{2\Omega_c^1}{N\Gamma}=1.74$ and $\frac{2\Omega_c}{N\Gamma}=2.125$. In the main panels, the orange lines are the exact numerical simulations, the dashed black lines are the mean field predictions and the dashed red lines are the exact steady state values. Even though the spontaneous emission rate sets a characteristic time scale of the order of  hundreds of seconds, one observes much faster dynamics due to the fact that $\Omega$ should scale proportionally to $ N$ to compete with collective emission and dissipation in the appropriate parts of the phase diagram. Thus, for typical experimental atom numbers  $N\approx 10^5$ one expects that most of the characteristic features (decay in A and B, and oscillations in D, E and F) can be observed  at $t<1s$.}\label{IIIInversion}
\end{figure}

In the metastable phase and for drive strengths much less than $\Omega_c^1$ (A in Fig.~\ref{IIIInversion}), mean field offers a good description of the dynamics, including the relaxation towards the steady state, which is shown to occur rather quickly compared to $\Gamma^{-1}\approx 1500$~s. The decay time predicted by mean field is $\sim 10$~s [Eq.~(\ref{IIIDecayEigenvalue})], which is faster than observed in Fig.~\ref{IIIInversion}. However, the discrepancy is explained by the fact that the system is initialized far from the steady state.

As $\Omega$ is increased, the mean field transition is reached (B and C in Fig.~\ref{IIIInversion}), which is signaled in the exact solution  by the fact that for $\Omega>\Omega_c^1$,  the inversion $\braket{\hat{J}_z}$ crosses  0 at some time during the evolution. Traces of oscillations can also be seen, but the qualitative behaviour of the exact solution is very similar in B and C (in contrast to the mean field predictions). Note, however, that the relaxation towards the steady state is very slow in this region (especially after crossing the boundary towards C).

\begin{figure}
\includegraphics[width=0.48\textwidth]{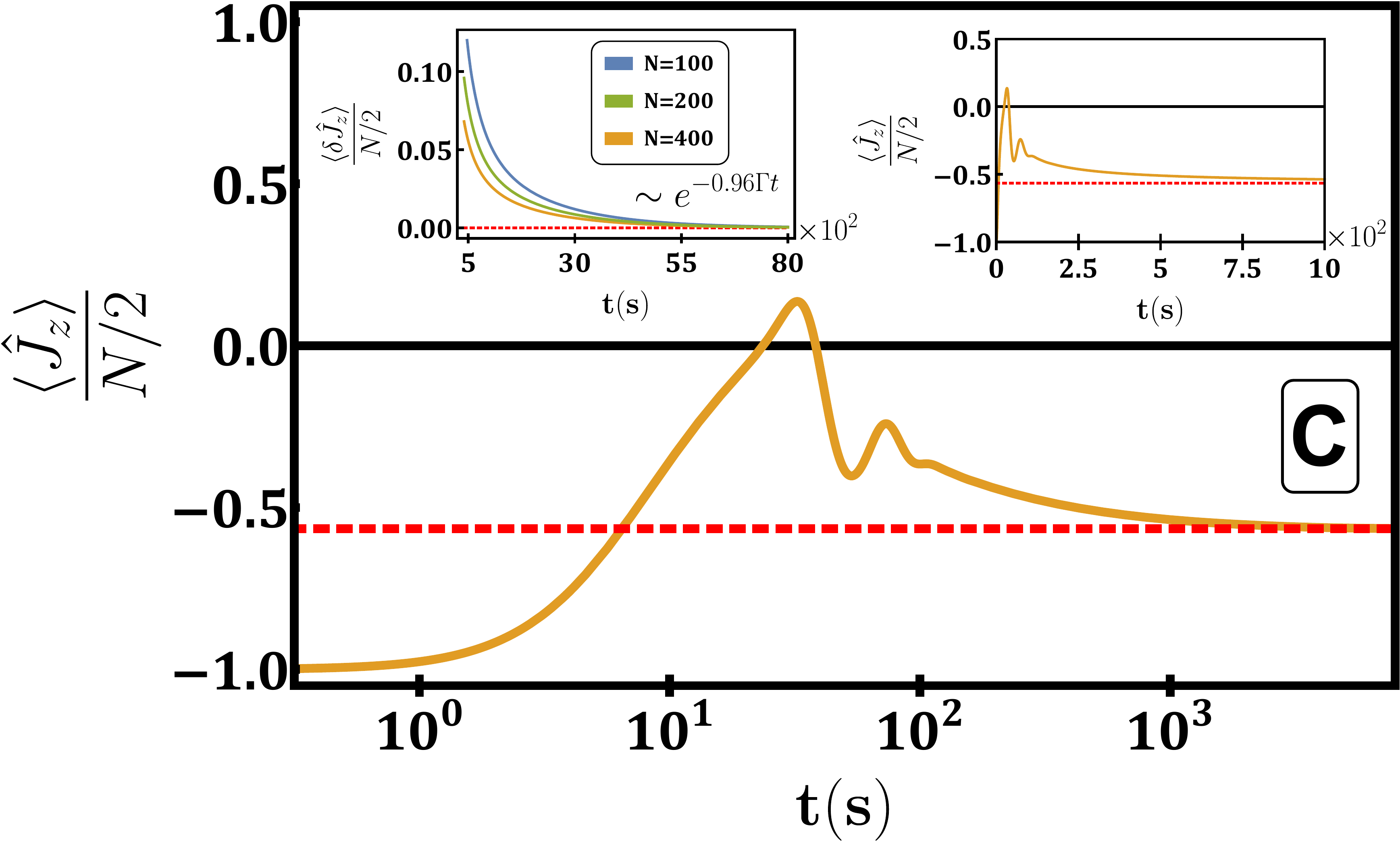}\hspace{0.6cm}
\includegraphics[width=0.48\textwidth]{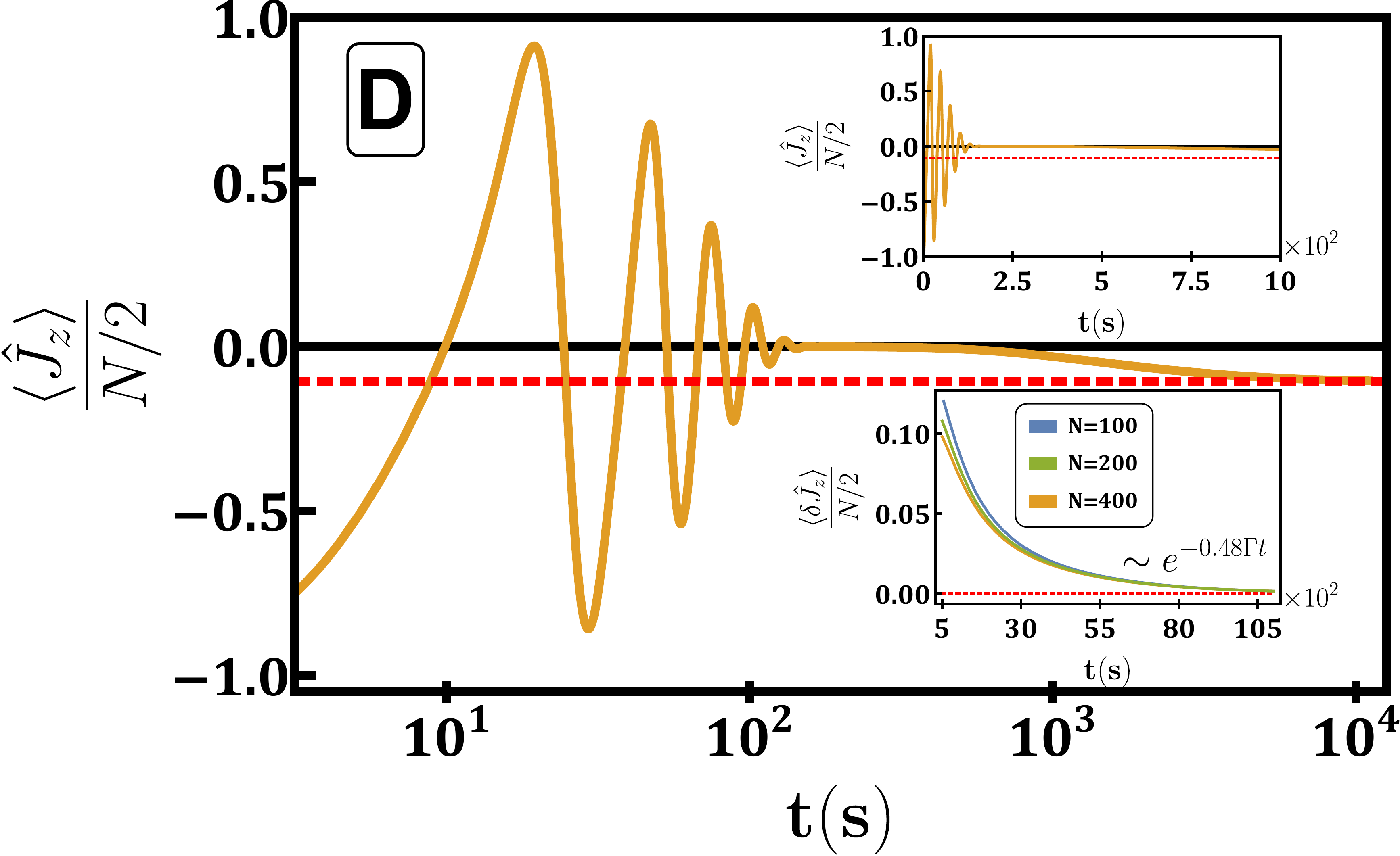}\caption{Simulations show that equilibration to the exact steady state in the metastable  region occurs at an extremely long timescales, of order $\sim \Gamma^{-1}$. 
Right inset in C and upper inset in D show the same plot but for a much shorter timespan. Left inset in C and lower inset in D compare the equilibration for different $N$ (with $\braket{\hat{J}_z}$ measured relative to its steady state), with the conclusion that $\braket{\hat{J}_z}$ approaches its steady state exponentially and in an $N$ independent way.}\label{IIILongTimeI}
\end{figure}

In the vicinity of $\Omega_c$ (D and E in Fig.~\ref{IIIInversion}), oscillations have already kicked in, and to a great extent the mean field solution  is able to capture the oscillation frequency (as can be seen by looking at the zero crossings). However, there is an envelope that describes relaxation towards $\braket{\hat{J}_z}=0$. This is the correct steady state for E since it is already in the $\mathcal N$ phase, but for D, which is still in the metastable  phase the real steady state is attained after a much longer time (set by $\Gamma$, see Fig.~\ref{IIILongTimeI}, right panel).

Finally, deep in the normal  phase (F in Fig.~\ref{IIIInversion}), the picture of oscillations modulated by an envelope is accurate. Note that it seems to be independent of $\Omega$ and thus the effect of increasing $\Omega$ is just to increase the oscillation frequency compared to its characteristic  decay rate. This envelope is a beyond mean field effect and, as shown in Appendix A, is characterized by $t_{\mathrm{decay}}\sim N^{-2/3}$.

Relaxation in C and D is very slow, so we extend the simulations to  long times to be able to observe the decay. The results are shown in Fig.~\ref{IIILongTimeI}. The decay is still exponential, but, in contrast to the superradiant region, there is no $N$-fold enhancement. The decay constants are almost independent of $N$, giving an interesting separation of scales for $\braket{\hat{J}_z}$, especially in point $D$: oscillations occur on a timescale of $O(N^{-1})$ (this is the mean field prediction), modulated by an envelope that decays to 0 on a timescale $O(N^{-2/3})$ and then relaxes to its true steady state in a time $\sim O(1)$.

\begin{figure}
\includegraphics[width=0.5\textwidth]{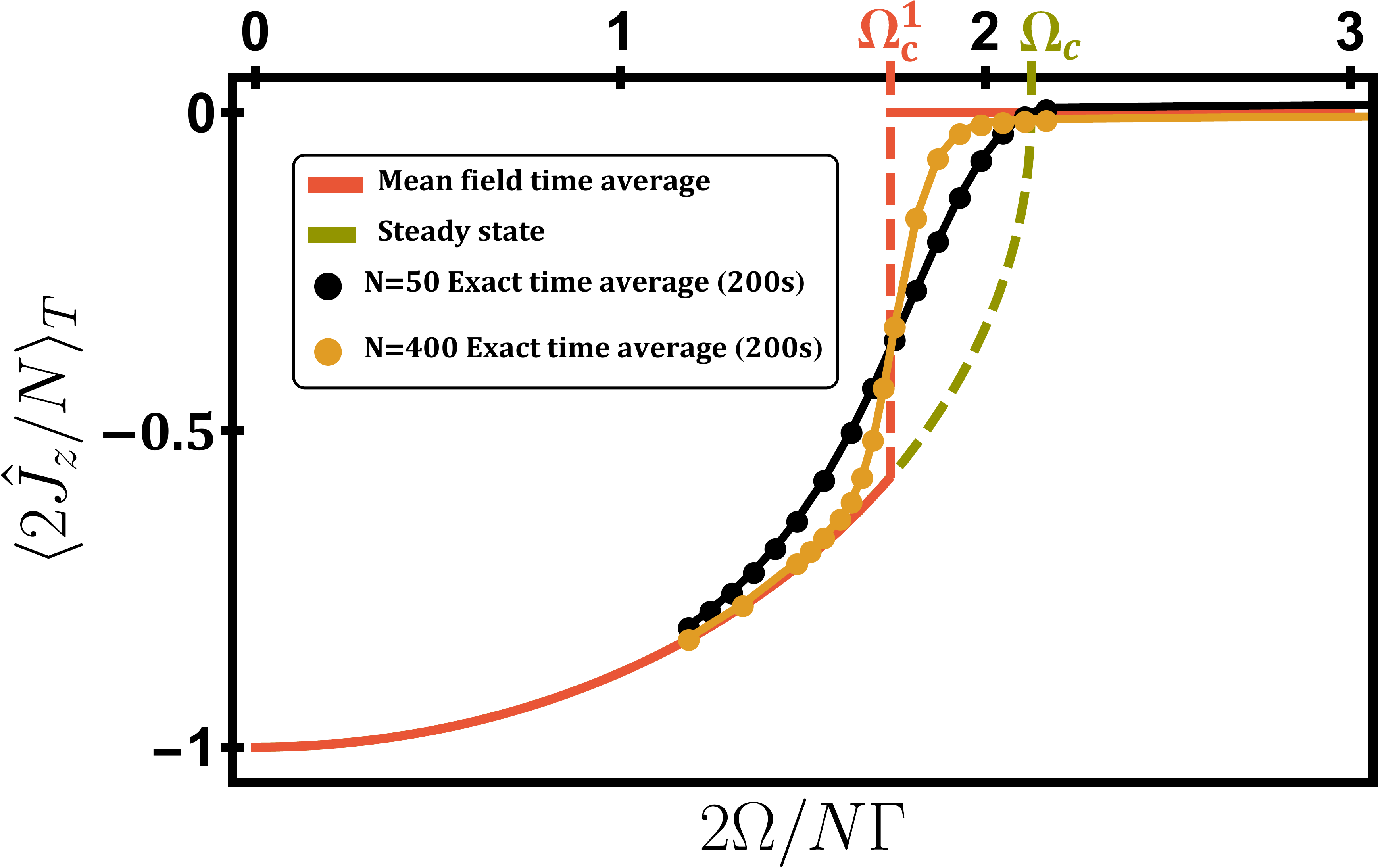}\caption{Simulations of the time averaged inversion during a $200$ s time window. Here we use $N=50$ and $400$. 
The exact dynamics  follow approximately the first order jump, but the transition is smoothed out by quantum fluctuations. The transition region seems to become narrower and the jump more pronounced as $N$ increases.}\label{IIIAverageValues}.
\end{figure}

We can also perform a time average. The time  window is chosen   to be short enough such that $D$ has not yet reached its true quantum steady state but long enough that it has relaxed to its metastable value. We take this characteristic time to be $\sim200$~s and plot in Fig.~\ref{IIIAverageValues} the corresponding time average value against the prediction of the mean field and exact steady state (see Fig.~\ref{IIFirstOrderTransition}). In contrast to the first order transition  observed at the mean field level, quantum fluctuations soften the sharp jump in the time average, especially at finite $N$. The sharp behaviour is only recovered in the thermodynamic limit $N\rightarrow\infty$.
\subsection{Squeezing dynamics}
\begin{figure}
\includegraphics[width=\textwidth]{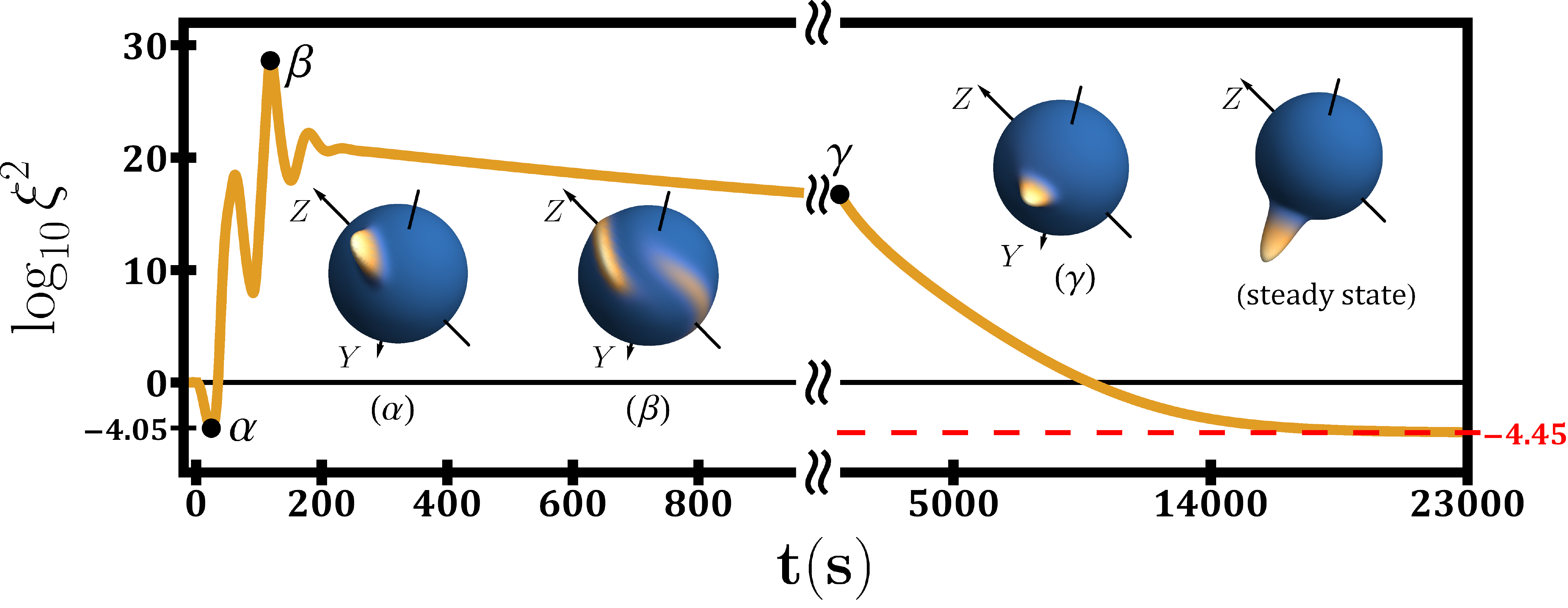}\caption{Exact spin squezing as a function of time for $N=200$, $\Gamma=6.7\times 10^{-4}\mathrm{s}^{-1}$, $\chi=6.3\times 10^{-4}\mathrm{s}^{-1}$ and $\Omega=0.938\Omega_c$. We chose this $\Omega$ because it gives the best squeezing for this number of particles. We show the numerical simulation (orange) and steady state squeezing (dashed red). We also show the Husimi distribution of the state at three points, marked $\alpha$, $\beta$ and $\gamma$ as well as the distribution of the steady state. We can clearly see that there is a dip at very short times ($\alpha$) beyond which the state becomes oversqueezed (e.g. $\beta$). As the system relaxes to equilibrium, beyond $\gamma$, the squeezing monotonically approaches its steady state value, though this occurs at very long times.}\label{IIIDynamicSqueezing}
\end{figure}

It is also worthwile to consider the dynamical behaviour of spin squeezing for two reasons: (1) To characterize the amount of spin squeeezing generated during transient dynamics,  and (2) to analyze how fast the steady state squeezing described in sec.~\ref{SSProperties} can be attained.
We show a plot of the spin squeezing as a function of time in Fig.~\ref{IIIDynamicSqueezing}, where we have chosen the value of $\Omega$ to be identical to the one that results in the largest steady state squeezing (the fact that it occurs at $\Omega\neq\Omega_c$ is a finite size effect). We also show the Husimi distribution of the state at specific points in time to better visualize the nature of the squeezing. At short times, we find a minimum, marked by $\alpha$ in Fig.~\ref{IIIDynamicSqueezing}, which rapidly evolves into an oversqueezed state, indicated by $\beta$. This is illustrated by the associated Husimi function for these points in Fig.~\ref{IIIDynamicSqueezing}: for $\alpha$, the state remains strongly polarized in one direction with a sheared noise ellipse,  whereas for $\beta$ the state wraps around the Bloch sphere. At later times, the distribution relocalizes on the Bloch sphere (point $\gamma$) and slowly relaxes to its steady state at much longer times (notice the timescale in Fig.~\ref{IIIDynamicSqueezing}).
\section{Spectrum of Liouville operator}\label{LOperator}
To better understand the different decay timescales and the onset of oscillatory behaviour, we study the spectrum of the Liouville operator, Eq.~(\ref{IMasterEquation}), and try to find signatures of the mean field transition in its eigenvalues. A gap in the real part of the spectrum translates into exponential decay of observables to their steady state. On the other hand, the imaginary part of the spectrum leads to nontrivial transient oscillatory dynamics.

\begin{figure}
\subfloat{\includegraphics[width=0.48\textwidth]{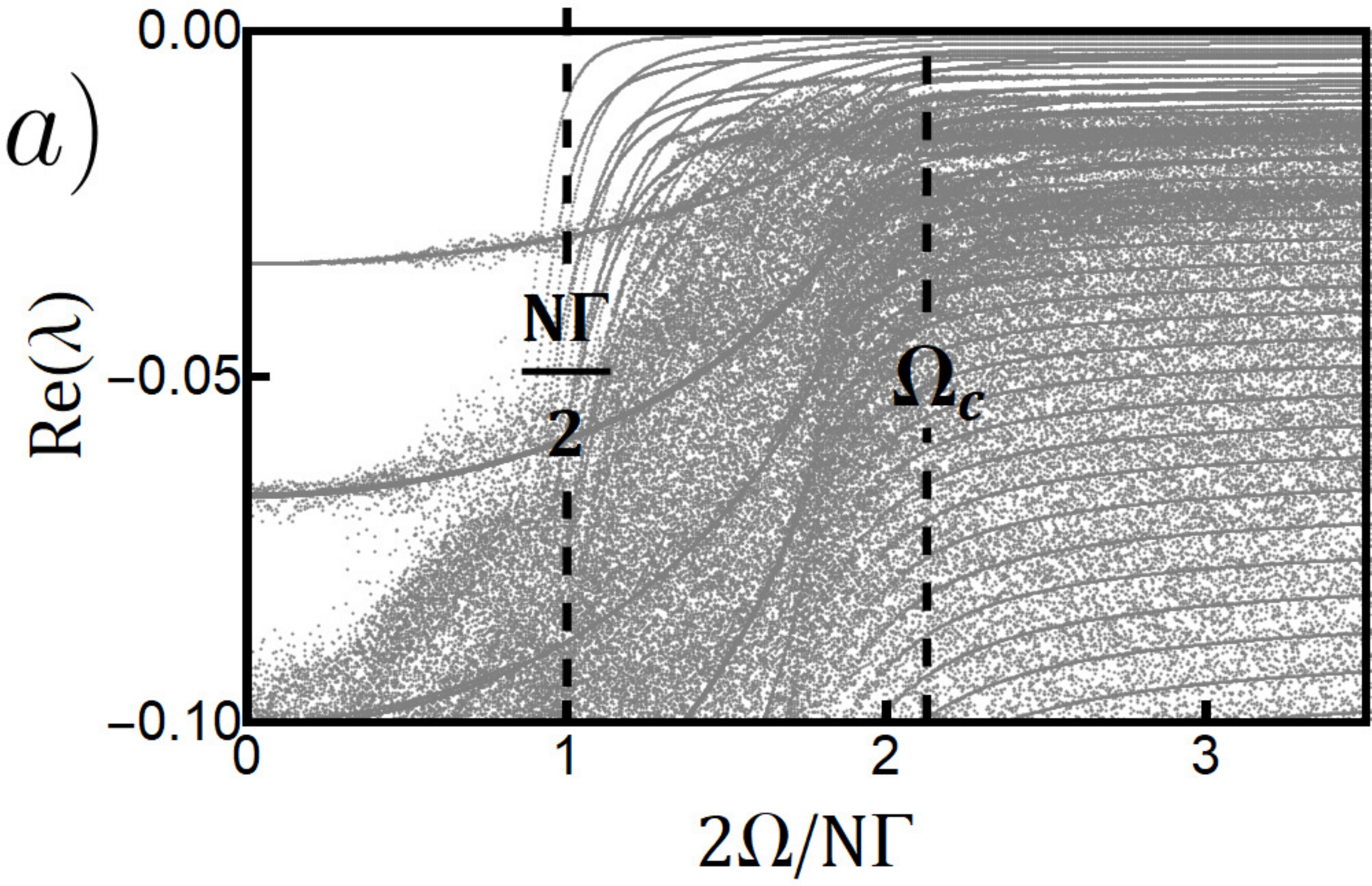}}\hspace{0.25cm}
\subfloat{\includegraphics[width=0.48\textwidth]{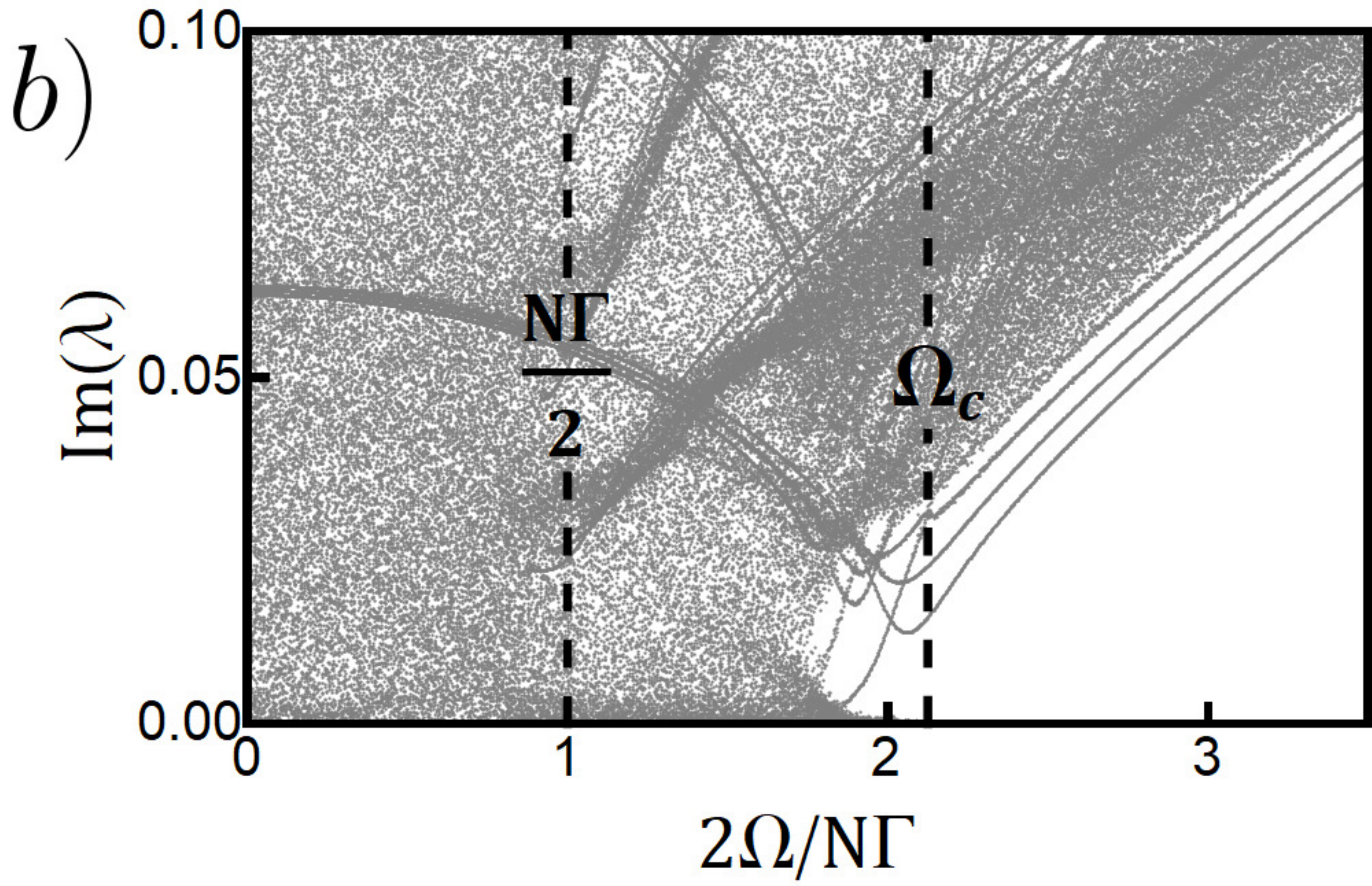}}
\caption{First 150 eigenvalues of the Liouville operator for $\frac{2\chi}{\Gamma}=1.875$ and $N=200$. a) The real part shows a gap for small $\Omega$. After a critical value, this gap closes and a continuum below 0 seems to arise in the thermodynamic limit. Note that the gap closes due to two eigenvalues crossing. b) The imaginary part shows the opposite behaviour. For small $\Omega$ the eigenvalues seem to fill a continuum (in the thermodynamic limit) but after another critical value, a gap opens in a seemingly continuous way.}\label{IVEigenvalue}
\end{figure}

For a system with $N=200$, in Fig.~\ref{IVEigenvalue} we plot the real and imaginary parts of the first 150 eigenvalues (ordered ascendingly with respect to their magnitude).
Very clearly, a gap closes in the real part at a value close to $ N\Gamma/2$, which implies that the decay times for $\Omega>N\Gamma/2$ can become very long. This is consistent with mean field since this is the region where oscillating nondecaying solutions begin to exist. The imaginary part of the eigenvalues remains gapless, however, indicating that nonoscillating behaviour is also allowed (metastable  region). Furthermore, a gap opens in the imaginary part close to $\Omega_c$, which implies that observables of the system should generally show oscillatory behaviour for $\Omega>\Omega_c$. Once again, this is consistent with the mean field analysis since in this region only oscillatory solutions exist. From the computed spectra,  we expect that, as $N$ increases, the gap  in the  real (imaginary) part should open(close) progressively closer to $ N\Gamma/2$ ($\Omega_c$) and in a more abrupt way.

\begin{figure}
\subfloat{\includegraphics[width=0.48\textwidth]{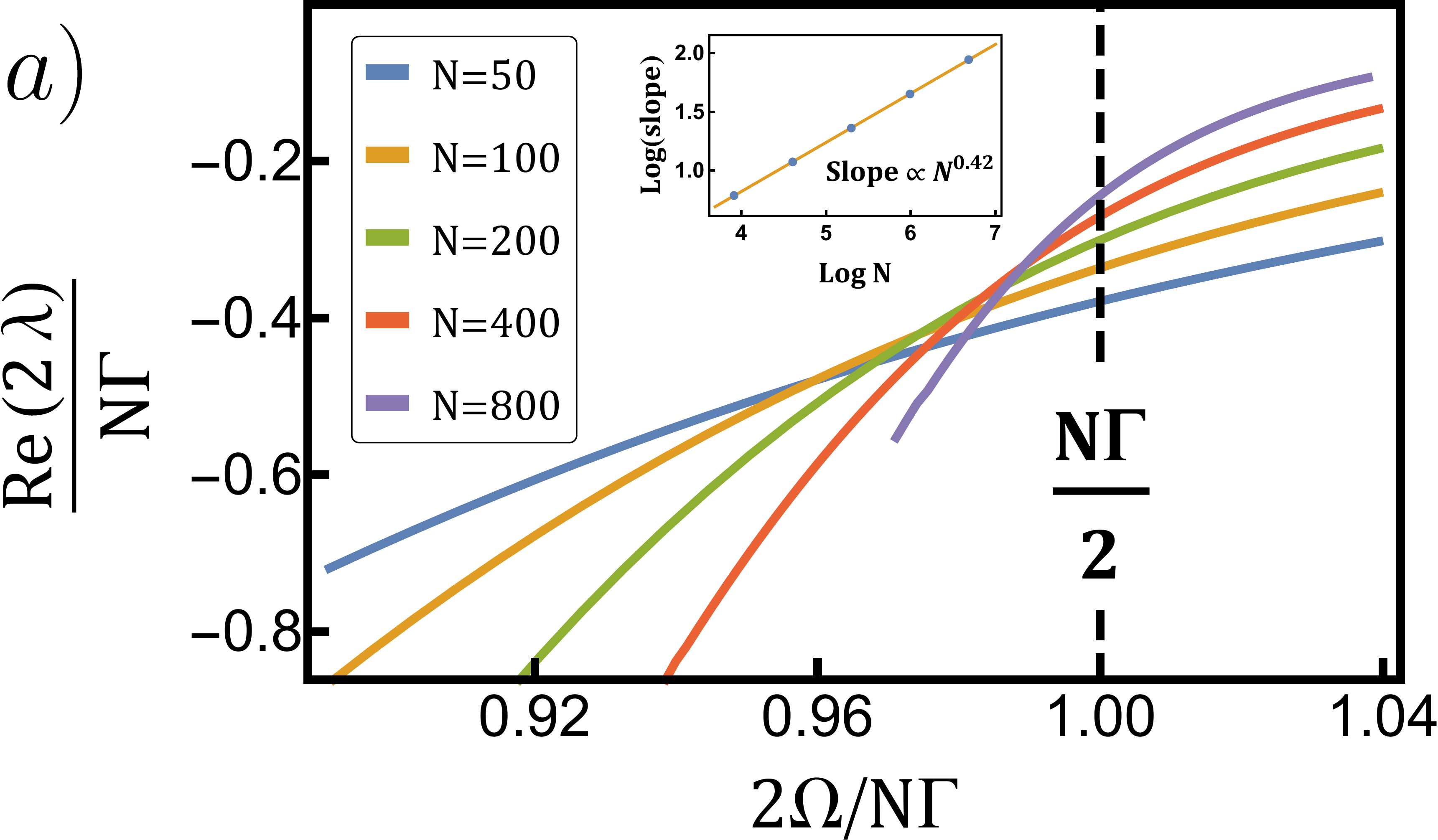}}\hspace{0.8cm}
\subfloat{\includegraphics[width=0.46\textwidth]{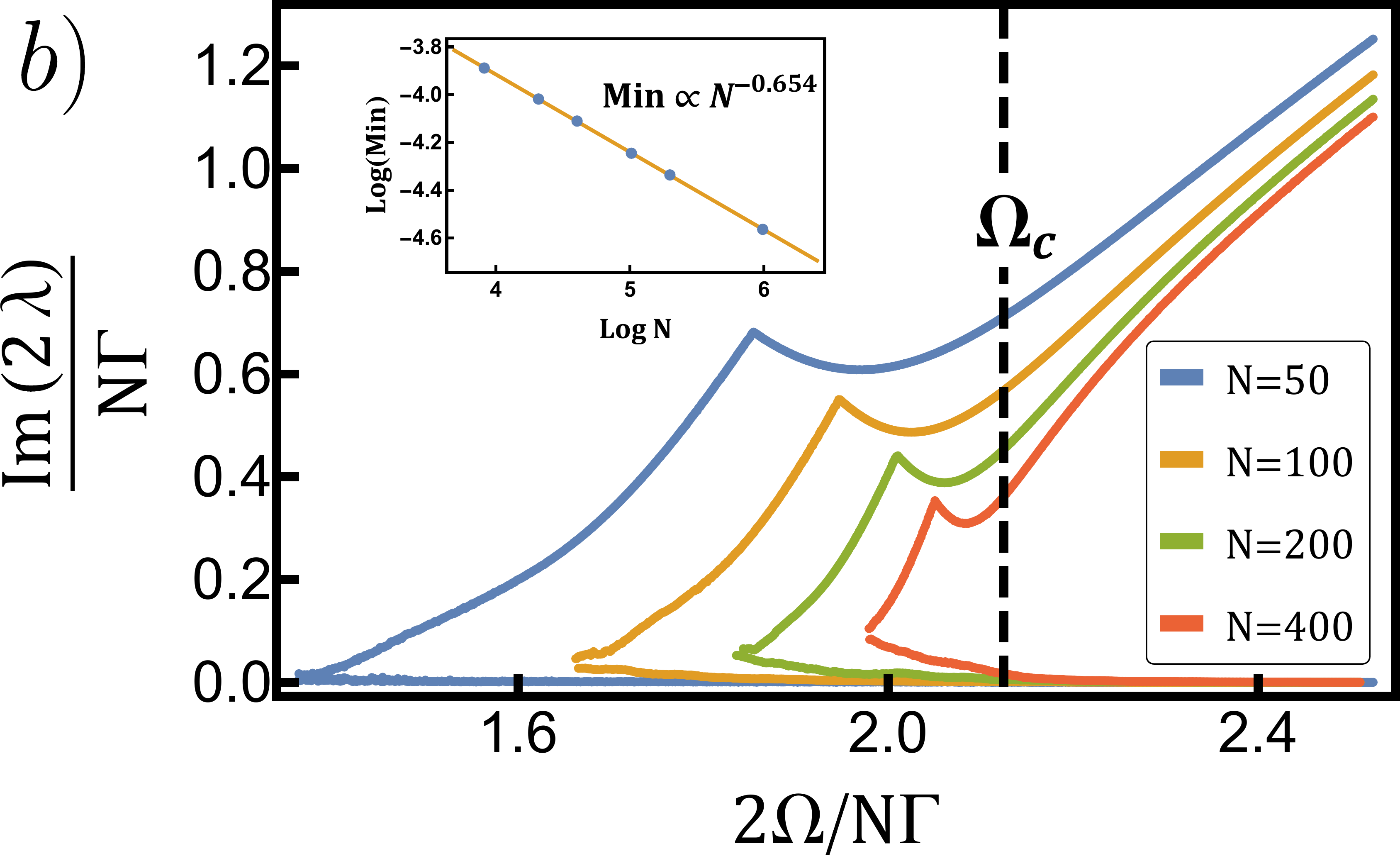}}
\caption{Left: Real part of the spectrum near the closing of the gap. It seems to close in a discontinuous fashion, with the slope of the transition region increasing as $N$ increases.  The inset shows the slope of the curve at $\Omega=\frac{1}{2}N\Gamma$, which seems to grow as a power law with $N$. Right: Gap in the imaginary part of the spectrum as a function of $N$. The cusp in the curve can be ascribed to the crossing of two eigenvalues (see Fig.~\ref{IVEigenvalue}). As $N$ increases, all the features of this opening (position of the cusp, local minimum to the right of the cusp, etc.),  seem to be approaching $\Omega_c$ (see inset).}\label{IVGap}
\end{figure}

In  Fig.~\ref{IVGap} we confirm this hypothesis. In panel (a) we show the region around which the gap in the real part of the spectrum closes. As $N\rightarrow\infty$, the curve becomes sharper and moves closer to $N\Gamma/2$. In panel (b), we show the gap in the imaginary part of the spectrum for different values of $N$ and observe  evidence that, effectively, the point at which it opens gets closer to $\Omega_c$ as $N\rightarrow\infty$ [see inset in panel (b)].

\section{Conclusions}
In this work, we have shown that the addition of interactions to a driven-dissipative collective system doesn't change qualitatively the steady state properties  but does affect significantly its dynamical behaviour in the limit of a large number of atoms. While in the noninteracting case both a slow relaxation and oscillations of the atomic inversion occur beyond a critical  driving strength, in the presence of interactions  a new metastable region appears. We  showed that this emergent behavior  could be understood both  via a mean field analysis, in which case the metastable region  manifests in the dynamics, and via an exact treatment,  where for a finite system fingerprints of  the metastable region can be observed in the spectrum of the Liouville operator. Finally, we also demonstrated that the slow relaxation present in one of the phases follows a scaling law different from the cases of pure interactions or pure dissipation, i.e. both must be present with comparably equal strengths to generate  a non-trivial relaxation dynamics.

Given recent progress in loading arrays of strontium atoms  in an optical cavity \cite{Norcia0}, and the capability to probe the associated electronic clock transition $^1S_0\rightarrow{}^3P_0$  we believe experiments are reaching the level of control to  observe the predicted behavior.
Nevertheless given the ultra-long relaxation  time in some of the phases it is  worthwhile to re-examine  in  the future  our predictions including possible extra decoherence processes  such as light scattering and atom loss, which are unavoidable in real experiments.

 An interesting parallel direction for future work would be to  use the  $^1S_0\rightarrow{}^3P_1$ transitions, which is still quite narrow compared to other dipole allowed transition (lifetime of the order of $20$~$\mu$s) but at the same time six orders of magnitude faster that the clock transition. The faster dynamics relaxes the stringent requirements on atom loss and decoherence on one hand, but on the other might invalidate the adiabatic elimination of the cavity photons. This is due to the comparable decay rates of the atomic transition and cavity. The necessity of including the photons during the dynamics might nevertheless make the dynamical phases richer and the phase diagram more complex than the one studied here, thus opening interesting new avenues of research.
\section{Acknowledgements} 
We thank Juan A. Muniz Silva, Zhe-Xuan Gong and Itamar Kimchi for feedback on the manuscript and Julia Cline for interesting discussions. This work is supported by the Air Force Office of Scientific Research grants FA9550-18-1-0319 and its Multidisciplinary University Research Initiative grant(MURI), by the Defense Advanced Research Projects Agency (DARPA) and Army Research Office grant W911NF-16-1-0576, the National Science Foundation grant PHY-1820885, JILA-NSF grant PFC-173400, and the National Institute of Standards and Technology.
\bibliographystyle{apsrev4-1.bst}
\bibliography{refs2.bib}
\appendix
\section{Envelope of oscillations in N phase}
\begin{figure}
\subfloat[Envelope and exact solution]{\includegraphics[width=0.46\textwidth]{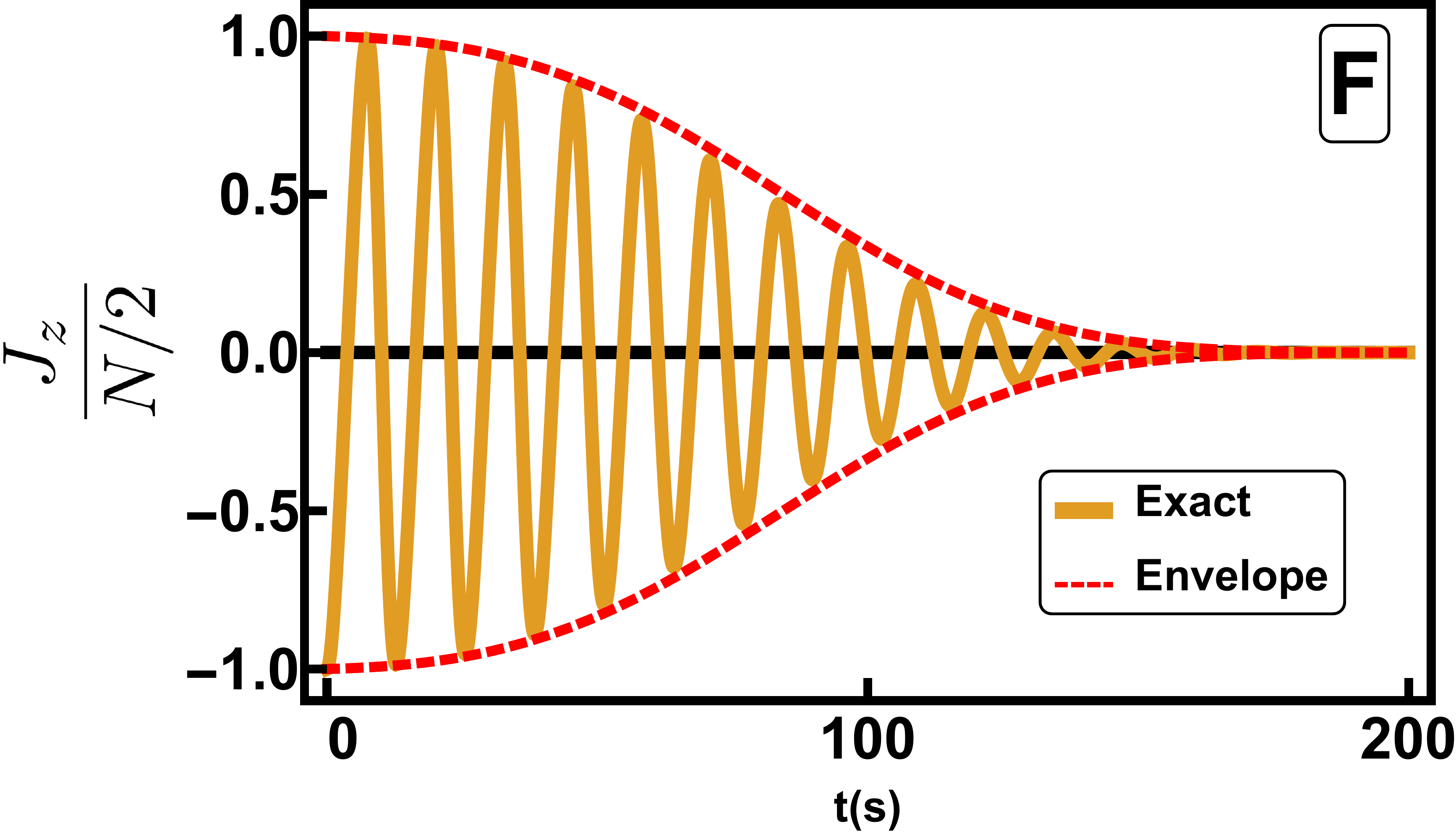}}\hspace{1cm}
\subfloat[Envelope scaling]{\includegraphics[width=0.47\textwidth]{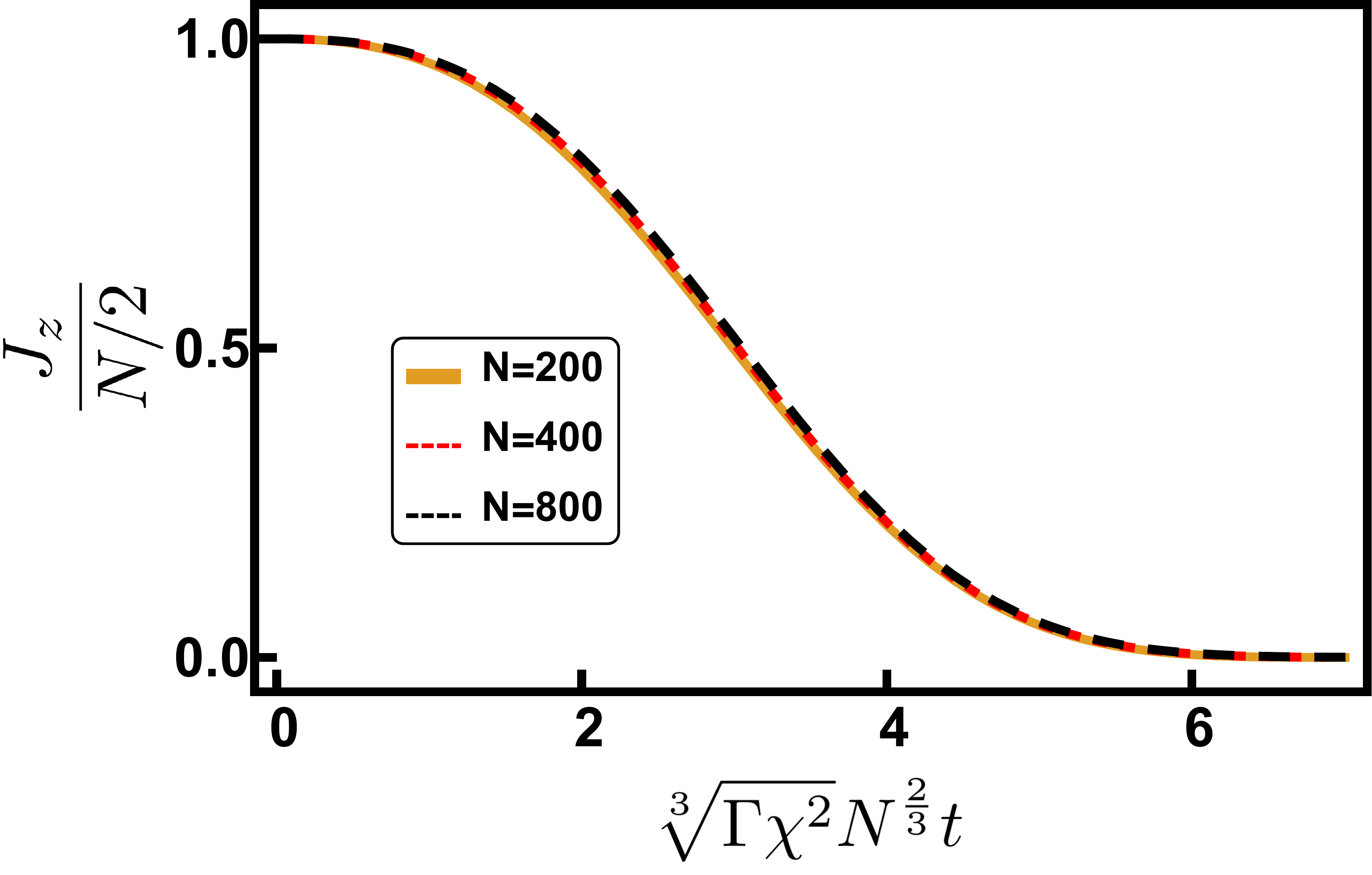}}
\caption{(a) Simulation of the point F in Fig.~\ref{IIIInversion} and of the envelope described by Eq.~(\ref{IIIEnvelopeME}) for $N=400$. There is almost perfect agreement. (b) Simulation of Eq.~(\ref{IIIEnvelopeME}) for $N=200$, $400$ and $800$. There are some subtle differences, especially at early times, but they become smaller the larger $N$ is.}\label{IIIEnvelope}
\end{figure}

\begin{figure}
\includegraphics[width=0.48\textwidth]{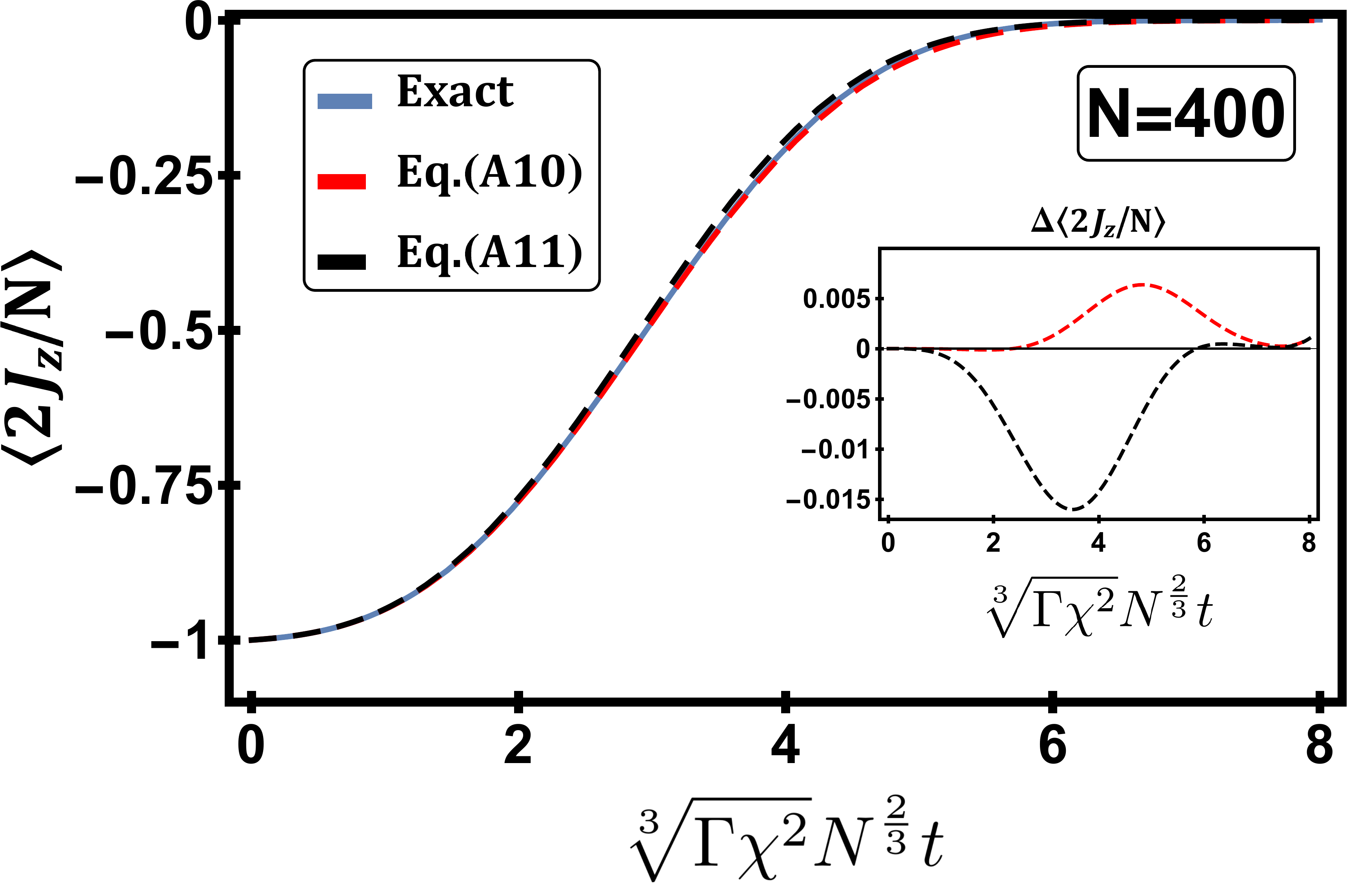}\hspace{0.5cm}
\includegraphics[width=0.48\textwidth]{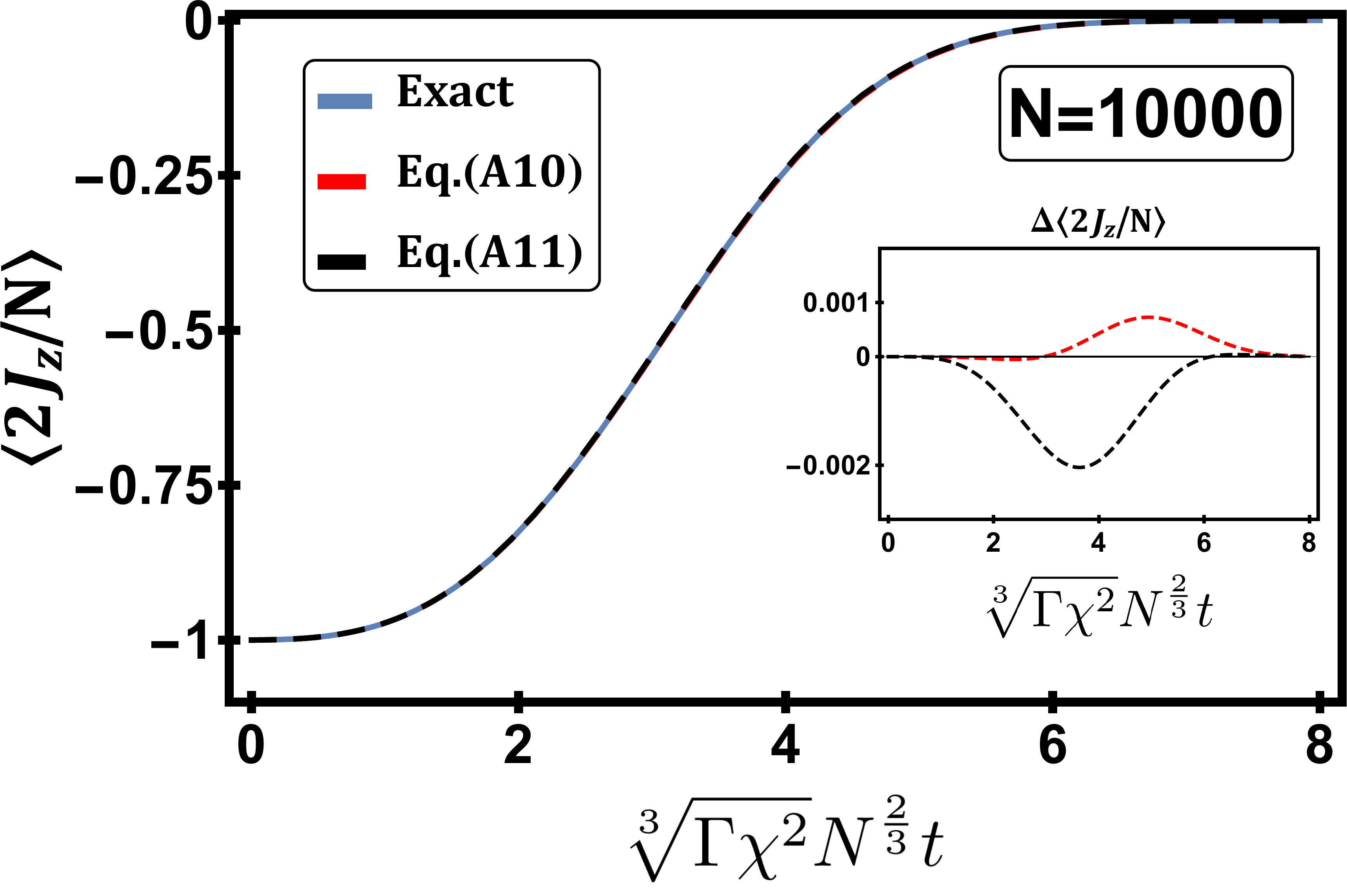}
\caption{We plot the exact envelope against the approximations given by Eqs.~(\ref{EnvelopeSol1}) and~(\ref{EnvelopeSol2}). The left panel corresponds to $N=400$ and the right one to $N=10000$. The insets show the difference between both approximations and the exact solution.}\label{EnvelopeApprox}
\end{figure}

To understand the envelope in Fig.~\ref{IIIInversion}(F), we go into the interaction picture generated by the driving term, $\Omega \hat{J}_x$, and time average the resulting Liouville superoperator over one oscillation. The ensuing equation for the time averaged density matrix, $\hat{\rho}^E$, is:
\begin{equation}\label{IIIEnvelopeME}
\frac{d\hat{\rho}^E}{dt}=-i\frac{\chi}{2}\big[\hat{J}_x^2,\hat{\rho}^E\big]+\frac{\Gamma}{2}\big[\hat{J}_x,\big[\hat{\rho}^E,\hat{J}_x\big]\big]+\frac{\Gamma}{4}\big[\hat{J}_z,\big[\hat{\rho}^E,\hat{J}_z\big]\big]+\frac{\Gamma}{4}\big[\hat{J}_y,\big[\hat{\rho}^E,\hat{J}_y\big]\big],
\end{equation}
and, as shown in Fig.~\ref{IIIEnvelope}(a), describes correctly the envelope of $\hat{J}_z$. The situation for $\hat{J}_x$ and $\hat{J}_y$ is worse, however, because for finite $\Omega$ they have a nonzero steady state value. To describe them accurately, higher order terms in the averaging expansion are needed (which is effectively an expansion in $\frac{1}{\Omega}$). The fact that Eq.~(\ref{IIIEnvelopeME}) is independent of $\Omega$ translates directly into the previously mentioned fact that the envelope is $\Omega$ independent.
It would be useful to understand how does the envelope scale with $N$, but even when working with Eq.~(\ref{IIIEnvelopeME}), there is no closed form solution. The decay of $\hat{J}_z$ is given by:
\begin{equation}	
\braket{\hat{J}_z}=-\frac{N}{2}e^{-\frac{3\Gamma t}{ 4}},
\end{equation}
in the case of $\chi=0$  with $t_{decay}\sim\Gamma^{-1}$ and
\begin{equation}
\braket{\hat{J}_z}=-\frac{N}{2}\bigg(\cos\frac{\chi t}{2}\bigg)^{N-1}\approx-\frac{N}{2}e^{-\frac{N\chi^2 t^2}{8}},
\end{equation}
for $\Gamma=0$ with $ t_{\mathrm{decay}}\sim \chi^{-1}N^{-\frac{1}{2}}$. To determine the scaling when both terms are present and $\chi\sim\Gamma$ (as is our case), we plotted the envelope for $N=200$, $400$, $800$, extracted the $e^{-\frac{3\Gamma t}{ 4}}$ time dependence (since it always dominates at short times) and then scaled the graphs to put them on top of each other, as shown in Fig.~\ref{IIIEnvelope}(b). The scaling exponent found was $t_{\mathrm{decay}}\sim N^{-0.66}$, which is corroborated by perturbation theory calculations, which indicate a leading dependence of the form of Eq.~(\ref{IIIEnvelopeF}) to sixth order. The exponent is then predicted to be $\frac{2}{3}$, very much in accordance with the numerical result,
\begin{equation}\label{IIIEnvelopeF}
-\frac{N}{2}e^{-\frac{\Gamma\chi^2}{48}N^2t^3}.
\end{equation}
A more satisfactory description of this behaviour can be obtained in the following way: we multiply Eq.~(\ref{IIIEnvelopeME}) by $\tilde{J}^+=\hat{J}_z-i\hat{J}_y$ on the right and then do a projection to obtain the equation satisfied by the matrix element $\bra{m}\rho\tilde{J}^+\ket{m}\equiv f(m)$, where $\ket{m}$  are eigenstates of $\hat{J}_x$.
\begin{equation}\label{AppME}
\frac{\partial f(m)}{\partial t}=-\frac{\Gamma}{4}f(m)+\frac{i\chi(2m+1)}{2}f(m)+\frac{\Gamma}{4}\bigg[\frac{N}{2}\bigg(\frac{N}{2}+1\bigg)-m(m+1)\bigg]\bigg[f(m+1)+f(m-1)-2f(m)\bigg].
\end{equation}
Since $f(t=0,m)\approx\sqrt{\frac{N}{2\pi}}e^{-\frac{m^2}{N/2}}$ we introduce a variable $x=\frac{m}{\sqrt{N/2}}$, which becomes continuous in the limit $N\rightarrow\infty$, and define the small quantity $\Delta=\frac{1}{\sqrt{N/2}}$. In terms of these, Eq.~(\ref{AppME}) takes the more sugestive form of Eq.~(\ref{AppME2}).
\begin{equation}\label{AppME2}
\frac{\partial f(x)}{\partial t}=-\frac{\Gamma}{4}f(x)+i\chi\bigg(\sqrt{\frac{N}{2}}x+\frac{1}{2}\bigg)f(x)+\frac{\Gamma}{4\Delta^4}\bigg(1-\Delta^2x^2+\Delta^2-x\Delta^3\bigg)\bigg[f(x+\Delta)+f(x-\Delta)-2f(x)\bigg],
\end{equation}
and we are now considering $f$ as a function of $x$ too. The limit $N\rightarrow\infty$ corresponds to $\Delta\rightarrow0$ upon which we obtain a partial differential equation (omitting terms of order $\Delta$),
\begin{equation}\label{AppME3}
\frac{\partial f}{\partial t}=-\frac{\Gamma}{4}f+i\chi\bigg(\sqrt{\frac{N}{2}}x+\frac{1}{2}\bigg)f+\frac{N\Gamma}{8}\frac{\partial^2f}{\partial x^2}+\frac{\Gamma}{4}(1-x^2)\frac{\partial^2f}{\partial x^2}+\frac{\Gamma}{48}\frac{\partial^4f}{\partial x^4}.
\end{equation}
In principle, we would have to solve this equation with the initial condition $f(0,x)=\sqrt{\frac{N}{2\pi}}e^{-x^2}$, corresponding to a spin pointing in the $z$ direction, and then integrate with respect to $x$ to obtain the average value of $\braket{\hat{J}_z-i\hat{J}_y}$ at later times, since:
\begin{equation}
\braket{\tilde{J}^+}=\Tr(\rho \tilde{J}^+)=\sum_m\bra{m}\rho\tilde{J}^+\ket{m}=\sum_mf(m)=\sqrt{\frac{N}{2}}\int f(x)dx.
\end{equation}
This becomes easier to treat in Fourier space (with respect to $x$), in which case the relevant equations are:
\begin{align}
\begin{split}
\tilde{f}(0,k)&=\sqrt{\frac{N}{2}}e^{-\frac{k^2}{4}},\\
\braket{\tilde{J}^+}&=\sqrt{\frac{N}{2}}\tilde{f}(t,0),\\
\frac{\partial \tilde{f}}{\partial t}&=\bigg(-\frac{3\Gamma}{4}+\frac{i\chi}{2}-\frac{N\Gamma k^2}{8}-\frac{\Gamma k^2}{4}+\frac{\Gamma k^4}{48}\bigg)\tilde{f}-\bigg(\frac{N\chi}{2}+\Gamma{k}\bigg)\frac{\partial \tilde{f}}{\partial k}-\frac{\Gamma k ^2}{4}\frac{\partial^2 \tilde{f}}{\partial k^2}.
\end{split}
\end{align}
If we neglect the second order partial derivative (since $\tilde{f}(0,k)$ is slowly varying) a closed form expression can be written for the solution (we also neglect the $k^4$ term for simplicity),
\begin{equation}\label{EnvelopeSol1}
\braket{\hat{J}_z}=\frac{N}{2}\exp\bigg[{-\frac{3\Gamma t}{4}-\frac{N\chi^2(1-e^{-\Gamma t})^2}{8\Gamma^2}}+\frac{N(N+2)\chi^2(e^{-2\Gamma t}-4e^{-\Gamma t}+3-2\Gamma t)}{32\Gamma^2}\bigg]\cos{\frac{\chi t}{2}}.
\end{equation}
For $\chi t,\Gamma t\ll 1$, this becomes
\begin{equation}\label{EnvelopeSol2}
\braket{\hat{J}_z}=\frac{N}{2}\exp\bigg[-\frac{3\Gamma t}{4}-\frac{\chi^2Nt^2}{8}-\frac{\chi^2\Gamma N(N+2)t^3}{48}\bigg],
\end{equation}
which agrees with the results from perturbation theory, up to to an irrelevant minus sign. We can compare both expressions, Eqs.~(\ref{EnvelopeSol1}) and~(\ref{EnvelopeSol2}), to the exact solution for $N=400$ and $N=10000$ (Fig.~\ref{EnvelopeApprox}). Finally, we remark that this formulation of long time corrections to mean field using a single partial differential equation is always achievable after doing an averaging over a single particle constant drive, as in the case of Eq.~(\ref{IIIEnvelopeME}). If we tried to do this with the original master equation, we would find, instead, an infinite set of coupled partial differential equations.

\newpage

\end{document}